\RequirePackage{fix-cm}
\pdfoutput=1
\documentclass[spbasic,smallextended]{svjour3}       

\smartqed  
\usepackage{graphicx}
\usepackage{cite,amssymb,amsmath,amsfonts,graphicx,bm,dcolumn,mathrsfs}
\usepackage{amsmath,amssymb,amsfonts,graphicx,bm,dcolumn,mathrsfs}
\usepackage{color}
\usepackage{latexsym}
\usepackage{dcolumn}
\usepackage{bm}
\usepackage[normalem]{ulem}
\usepackage{mathtools}

\newcommand{\be}{\begin{equation}}
\newcommand{\ee}{\end{equation}}
\newcommand{\bea}{\begin{eqnarray}}
\newcommand{\eea}{\end{eqnarray}}
\newcommand{\bse}{\begin{subequations}}
\newcommand{\ese}{\end{subequations}}
\newcommand{\p}{\partial}

\newcommand{\e}{\epsilon} 

\newcommand{\comment}[1]{}

\renewcommand{\ss}[1]{_{\hbox{\tiny #1}}}

\begin{document}

\title{Localization in boundary-driven lattice models}
\author{Michele Giusfredi$^{1,2,3}$, Stefano Iubini$^{2,3}$ and Paolo Politi$^{2,3}$}
\institute{Michele Giusfredi \at Michele.Giusfredi@unifi.it
           \and Stefano Iubini \at Stefano.Iubini@cnr.it
	\and Paolo Politi \at Paolo.Politi@cnr.it 
\and
              $^1$ Dipartimento di Fisica e Astronomia, Universit\`a di Firenze,
via G. Sansone 1 I-50019, Sesto Fiorentino, Italy \at
\and
$^2$ Istituto dei Sistemi Complessi, Consiglio Nazionale
delle Ricerche, via Madonna del Piano 10, I-50019 Sesto Fiorentino, Italy \at
\and
$^3$ Istituto Nazionale di Fisica Nucleare, Sezione di Firenze,
via G. Sansone 1 I-50019, Sesto Fiorentino, Italy
}
\date{Received: \today}
\authorrunning{M. Giusfredi, S. Iubini, P. Politi}
\titlerunning{Localization in boundary-driven lattice models}
\maketitle

\begin{abstract}
Several systems display an equilibrium condensation transition,
where a finite fraction of a conserved quantity is spatially
localized. The presence of two conservation laws may induce
the emergence of such transition in an out-of-equilibrium setup,
where boundaries are attached to different and subcritical
heat baths. We study this phenomenon in a class of stochastic lattice models, 
where the local energy is a general convex function of the local mass,
mass and energy being both globally conserved in the isolated system.
We obtain exact results for the nonequilibrium steady state (spatial profiles, mass and energy currents, Onsager coefficients)
and we highlight important differences between equilibrium and
out-of-equilibrium condensation.

\end{abstract}

\section{Introduction}
\label{sec.intro}

Condensation transitions have been extensively studied in the last decades and they can correspond to rather different processes:
they can be understood as localization processes where a macroscopic fraction of a conserved quantity is concentrated in a finite region of space, but they might also indicate a large deviation phenomenon where a single random variable gives a finite contribution to the constrained sum of many variables. From a physical point of view, condensation processes can be related to phenomena ranging from aggregation/fragmentation~\cite{eggers99,aggregation_condensation} to localization in propagating light~\cite{Eisenberg1998,russell2023localization}, from discrete solitons in Bose-Einstein condensates~\cite{TS} to the formation of jamming in driven flows~\cite{traffic_flow_review,soh2018jamming}, from wealth condensation in social economy~\cite{Burda_wealth} to localization phenomena in networks~\cite{gelation2000,Pastor2016_SR}.

\comment{
The existence of equilibrium condensation transitions in one dimensional systems is not in contradiction with the Landau-Peierls argument~\cite{Landau_book} or with the Perron-Frobenius theorem~\cite{Meyer_book}: in fact, it turns out that lattice models displaying condensation require the use of a transfer matrix of divergent size, which may reveal the existence of effective long-range interactions; a feature, this, which is well known to allow for one-dimensional phase transitions and also to be at the origin of inequivalence between statistical ensembles~\cite{Ruffo_inequivalence}.
}
Conservation laws are known to play  a crucial role for equilibrium condensation transitions. As an example, a class of condensation models which has played a relevant role in the advancement of this field is that of mass-transfer models with a factorized steady state.\footnote{In many cases the system is at/out-of equilibrium according to the symmetry/asymmetry of the transfer process, but the factorized state, therefore the nature of the phase transition, does not change.} Let us suppose that a conserved quantity $X=\sum_{i=1}^N x_i$ is shared among $N$ sites $i$, each $x_i$ being distributed according to some function $f(x)$.
If $f(x)$ decays slower than exponentially, a condensation process occurs as a cooperative effect above a certain critical value $X/N>x_c$. In~\cite{Szavits2014_PRL} it was shown that the presence of additional conservation laws can enforce condensation even  with light-tailed distributions, exhibiting exponentially or faster decays.  The opposite outcome can occur with fat-tailed distributions, i.e. the suppression of a condensed phase that was originally present with a smaller number of conservation laws.

\begin{figure}
        \includegraphics[width=0.8\columnwidth]{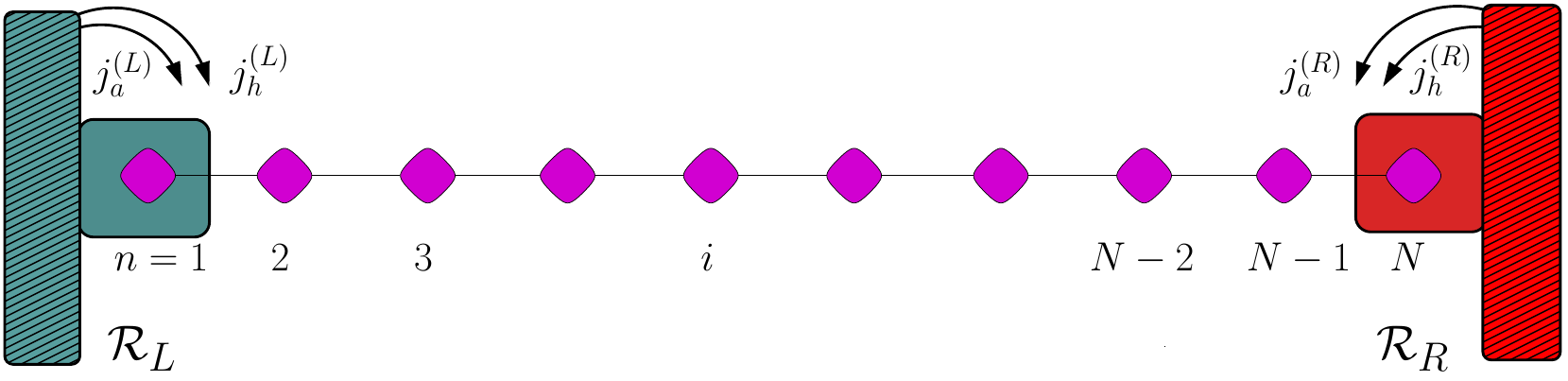} 
        \centering
        \caption{
        The non equilibrium setup: a chain of $N$ sites attached at its ends with two external reservoirs with thermodynamic parameters $\mathcal{R}_L$ and $\mathcal{R}_R$. Each reservoir exchanges mass 
        $(a)$ and energy $(h)$ with the chain, whose corresponding fluxes are denoted by $j_{a,h}^{(L,R)}$, see Sec. \ref{sec.baths}.
        }\label{fig:setup}
    \end{figure}

A much less explored field  concerns the study of 
condensation phenomena in steady out-of-equilibrium
conditions. 
Specifically, we refer to the typical transport setup in which two external reservoirs with  thermodynamic parameters $\mathcal{R}_L$ and $\mathcal{R}_R$ are in contact at the left and right boundaries of the system, respectively (see Fig.~\ref{fig:setup}).
The bulk dynamics is assumed to be local, reversible and constrained by a certain number $\nu\geq 1$ of independent conservation 
laws, while irreversibility is induced solely by boundary forces when $\mathcal{R}_L\neq\mathcal{R}_R$.
In this case, the resulting nonequilibrium stationary state (NESS) breaks time reversal and $\nu$ stationary conserved currents flow through the system. 
The occurrence of condensation in these conditions displays interesting features that crucially depend on the number of conservation laws.
In systems with $\nu=1$, condensation appears only if a boundary reservoir imposes overcritical conditions~\cite{Levine2005}.   
This fact can be easily understood because a locally conserved observable, $O$, 
should have a spatially monotonous profile $O(i)$
varying between the values $O(1) = O_L$ and $O(N)=O_R$ imposed by the reservoirs:
therefore, if $O_{L,R}$ are subcritical, the whole system is.

Recent studies of some of us for a specific model with $\nu=2$~\cite{gotti22,iubini23} 
have shown that in this case a condensed  state may arise in the bulk
of the system even in the presence of subcritical boundary conditions, 
provided that  $\mathcal{R}_L\neq\mathcal{R}_R$.\footnote{%
Previously, this possibility had appeared in a different out-of-equilibrium setup, where one of the two subcritical reservoirs  was replaced by a dissipator~\cite{Iubini2017_Entropy}.} 
The origin of this peculiar behavior can be traced back to the coupled transport between the two stationary currents 
and it is the result of a sort of extreme Joule effect.
For example, if a linear conductor is subject to a temperature gradient and to an electric potential gradient,
the heating induced by the Joule effect in the bulk of the sample may induce a non monotonous profile of the temperature.
In our case we have a coupled transport of mass and energy and the ``Joule" effect is \textit{extreme} because
the system can even attain overcritical conditions in the bulk
while the boundaries are maintained below the condensation threshold.

In Ref.~\cite{iubini23} the focus was the determination of the Onsager coefficients in the homogeneous phase,
which allowed to show that steady, not condensed profiles can be forced to enter the localization region.
In Ref.~\cite{gotti22} out-of-equilibrium condensation appeared from numerics.
In this work we make two important steps forward:
we provide a detailed analytical study of the nonequilibrium steady state
and we consider a broad class of condensation models with $\nu=2$, which includes
the model studied in ~\cite{gotti22,iubini23} but also other models
studied previously at equilibrium~\cite{Szavits2014_PRL,johansson04,Samuelsen2013},
therefore offering a unified vision of their properties.
This class of models and their equilibrium phase diagrams are defined and discussed
in Sec.~\ref{sec.equilibrium}, which therefore generalizes known results for specific models.	 

In Sec.~\ref{sec.baths} we discuss the out-of-equilibrium setup and the different types of reservoirs that can be employed.
This section allows the non specialist reader to gain confidence with the algorithms used throughout our work.

In Sec.~\ref{sec.ness} we present the core of our new results: the exact nonequilibrium spatial profiles of mass and energy,
Eqs.~(\ref{eq.aihi}), and the parametric profile, Eq.~(\ref{eq:lin_ha}), which is exactly linear.
All these results do not depend on the specific model in the class here considered and, importantly, hold also in the regime of nonequilibrium condensation.
%
Next, we provide an analytic expression for the currents, Eqs.~(\ref{eq.ja}-\ref{eq.jh}), and of the Onsager
coefficients $L_{ij}$, Eq.~(\ref{eq:Lxy}).
It is worth stressing that the local currents will be shown to display a ``universal" dependence on the mass and energy gradients,
while $L_{ij}$ are model-dependent due to their dependence on the gradients of the grandcanonical parameters, temperature and chemical potential.
 We also find an explicit expression for the Seebeck coefficient $\mathcal{S}$ close to the critical line and we show that $\mathcal{S}\neq 0$, thereby indicating the presence of thermodiffusive coupling in the sense
 of linear irreversible thermodynamics~\cite{benenti17rev}.

In Sec.~\ref{sec.differences} we discuss the differences between equilibrium and out-of-equilibrium condensation.
This part is supported by the simulation of a specific model, see Figs.~\ref{fig:en_pos_8k}-\ref{fig:Y2},
and by analytical considerations, valid for \textit{any} model, see Eq.~(\ref{eq.diffusion}) and below.
In practice, the nonequilibrium setup induces the birth and growth of peaks. If peaks can diffuse,
they are destined to die when they attain the subcritical regions: this is the dynamical process
at the origin of the NESS.

In Sec.~\ref{sec.discussion} we present a final discussion about nonequilibrium condensation and the role of a possible pinning of energy peaks.

\section{The models and their equilibrium properties}
\label{sec.equilibrium}

    \begin{figure}
        \includegraphics[width=1.0\columnwidth]{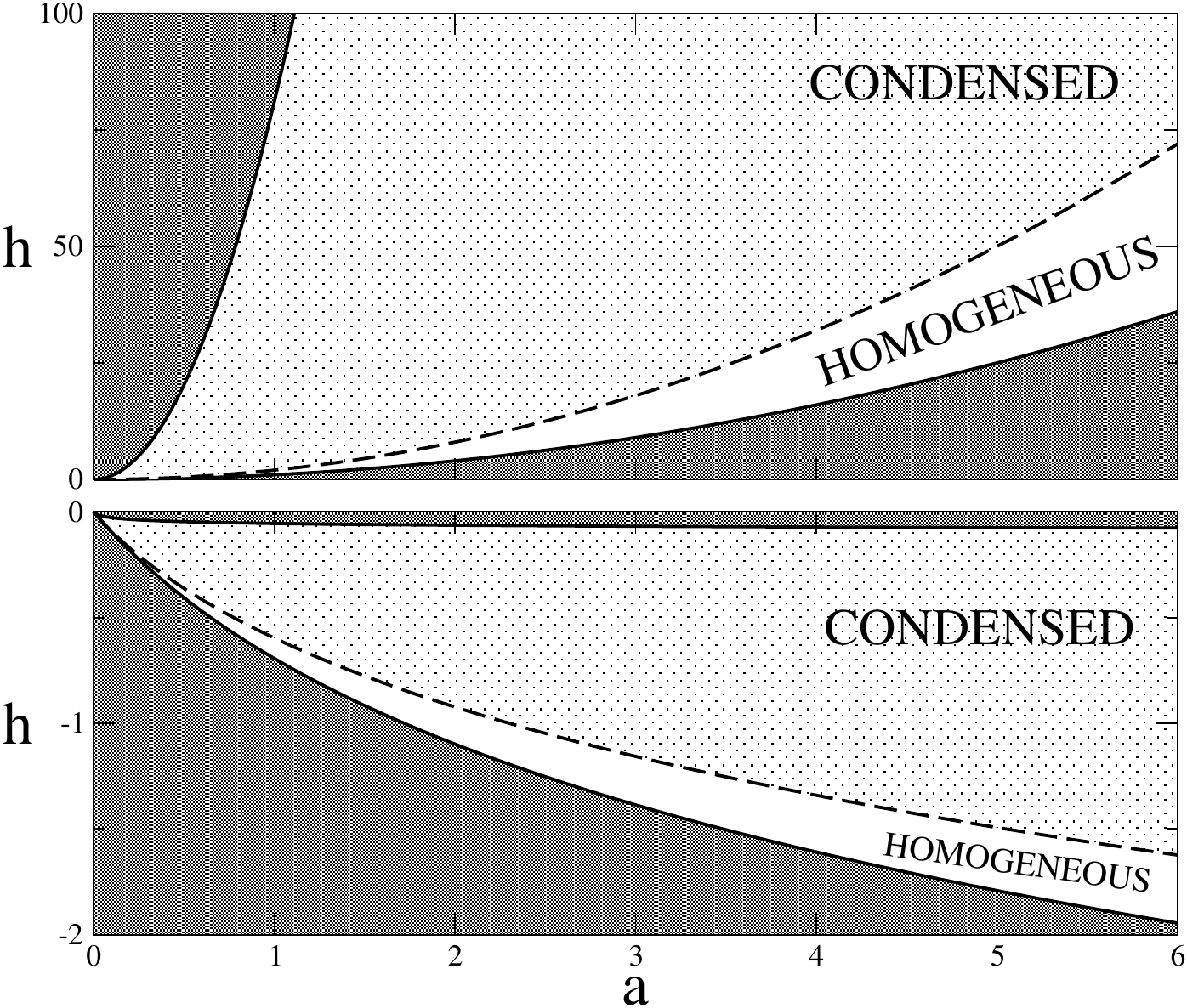} 
        \caption{Equilibrium phase diagrams for $F(c)=c^2$ (top panel) and for $F(c)=-\ln (1+c)$ (bottom panel). The grey regions are forbidden: $h<h\ss{GS}(a) = F(a)$ and $h>h\ss{M}(a)=N^{-1}F(Na)$ (for $N=81$). The dashed lines, $h\ss{C}(a)=\int_0^\infty dx F(ax)e^{-x}$ (see the main text), separate the homogeneous (white) phase from the condensed (dotted) phase. The latter cannot be described in the grandcanonical ensemble.
        }\label{fig.pd}
    \end{figure}

The class of models we are going to study is defined on 
a one-dimensional lattice of sites labeled by $i=1,\cdots, N$, where $N$ is the system size. On each site it
is defined a non-negative real variable $c_i\geq 0$, here called ``local mass'' and subject to a  stochastic evolution.
Another observable called ``local energy'' is defined through the relation $\e_i=F(c_i)$, where $F(c)$ is a convex function.
The total mass ${A}=\sum_i c_i\equiv Na$ and the total energy ${H}=\sum_i \epsilon_i\equiv Nh$ are exactly conserved. We can assume without loss of generality that $F(0)=0$.

Because of the convexity of $F(c)$, if $y_i$ are positive quantities such that $\sum_i y_i =1$, it must be
\be
F\left( \sum_i y_i c_i\right) \le \sum_i y_i F(c_i) .
\ee
Choosing $y_i =1/N$, we obtain $F(a)\le h$. Therefore, the region below $h\ss{GS}(a) = F(a)$ is forbidden. The curve $h=F(a)$ corresponds to masses $c_i$ that are all the same and it is the ground state (zero-temperature) curve.

Similarly, using the superadditivity property of a convex function vanishing at the origin, 
\be
\sum_i F(c_i) \le F\left( \sum_i c_i\right),
\ee
we find that $H\le F(A)$, i.e., $h\le h\ss{M}(a)\equiv\tfrac{1}{N}F(Na)$. 
For $F(c)=c^\alpha$ (with $\alpha>1$ so as to ensure convexity) this means $h\ss{M}(a) = N^{\alpha -1} h\ss{GS}(a)$,
a curve which flattens to the vertical axis in the limit $N\to\infty$.
For $F(c)=-\ln (1+c)$, $h\ss{M}(a) = - N^{-1} \ln (1+Na)$, which flattens to the horizontal axis. The curve $h\ss{M}(a)$ corresponds to all mass localized in a single site $k$, $c_i =A\delta_{i,k}$.

In the region between the curves $h\ss{GS}(a)$ and $h\ss{M}(a)$ there exists a third curve $h\ss{C}(a)$ separating the homogeneous from the condensed phase  and corresponding to an infinite, positive temperature. Here we limit to derive it in a simple and non-formal way, putting the details of the calculations in Appendix~\ref{app.mc}. In the grandcanonical ensemble the weight of the configuration $\{ c_i\}$ is
proportional to $e^{-\beta[H(c_i) -\mu A(c_i)]}$, where $\beta$ is the inverse temperature and $\mu$ the chemical potential. 
Moreover, the equilibrium measure factorizes in the product of single-particle distributions $\rho(c)= z(\mu,\beta)^{-1} e^{-\beta[F(c) -\mu c]}$,
where $z(\mu,\beta)=\int_0^\infty dc\, e^{-\beta[F(c) -\mu c]}$ is the (single particle) partition function.
The infinite-temperature limit is consistently obtained letting $\beta\to 0^+$  with $\beta\mu$ finite in order to enforce finite 
mass densities~\cite{Rasmussen2000_PRL}. In this limit, imposing $\langle c\rangle_\rho =a$, where $\langle \cdot \rangle_\rho$ is the average over the distribution $\rho$,
implies that the equilibrium distribution decays exponentially,  $\rho(c)=(1/a)e^{-c/a}$ and that $\beta\mu=-1/a$.  The corresponding average value of the energy is
\be
h\ss{C}(a) =\langle F(c)\rangle = \int_0^\infty dc F(c) \rho(c) 
=\int_0^\infty dx F(ax) e^{-x}.
\ee
It is straightforward to prove that
\be
h\ss{C}\big(ya_1 + (1-y)a_2\big) \le y h\ss{C}(a_1) + (1-y)h\ss{c}(a_2),
\ee
therefore proving that $h\ss{C}(a)$ is convex as well.
For $F(c)=c^\alpha$, we obtain $h\ss{C}(a) =\Gamma(\alpha+1)a^\alpha$;
for $F(c)=-\ln (1+c)$, we obtain $h\ss{C}(a)= - e^{1/a} E_1(1/a)$, where $E_1(x)=\int_x^\infty du\, e^{-u} /u$ is the exponential integral function.
In Fig.~\ref{fig.pd} we plot the phase diagrams: 
the upper panel refers to $F(c)=c^2$, corresponding to the case studied
in Refs.~\cite{gotti22,iubini23};
the lower panel refers to
 $F(c)=-\ln (1+c)$ and corresponds to the case studied in Ref.~\cite{Samuelsen2013}.
The critical curves $h\ss{C}(a)$ are plotted as dashed lines.
In the rest of the article we will focus on a positive $F(c)$, with special emphasis on $F(c)=c^\alpha$, a case 
studied in Refs.~\cite{Szavits2014_PRL,johansson04}.

A rigorous treatment of the grandcanonical ensemble, via Laplace transform and saddle point method, see Appendix~\ref{app.mc}, shows that a solution for equations~(\ref{eq:state}) exists only for real positive $\beta$ values, i.e.
for energy densities below the critical line $h\ss{C}(a)$.
For $h>h\ss{C}$ the saddle-point solution  breaks down, implying that the grand-canonical ensemble is not defined in this region. 
The microcanonical partition function $\Omega(a,h)$ can instead be estimated by means of large-deviations techniques~\cite{Majumdar2005_PRL,Evans2006_JSP,Majumdar2010_LesHouches}, which account for the fact
that typical configurations are here characterized by a macroscopic fraction of the total energy $h$ concentrated on a single lattice site.
In fact, the value $h=h\ss{C}$  identifies a phase transition from a homogeneous to a localized phase. 
More precisely, 
for  $h> h\ss{C}$ a localized phase
is developed, with a macroscopic amount of energy $(h-h\ss{C})N$ condensed on few (eventually one) lattice sites.
A perfect localization is attained only in the thermodynamic limit, where the equilibrium state consists of a single 
peak superposed to an extensive background lying on the critical line $h=h\ss{C}$ and whose mass is distributed exponentially.

Estimates of the microcanonical entropy $S(a,h)=\ln[\Omega(a,h)]$  in the condensed region~\cite{Gradenigo2021_JSTAT} (from now on we limit to $F(c)=c^\alpha$, see Eq.~\ref{eq:omega}) show that   for finite excess energies $h-h\ss{C}$,
\be
S(a,h)\simeq S_0(a) -[(h-h\ss{C})N]^{1/\alpha}\,.
\label{eq:LD_entr}
\ee
Here $S_0(a)$ is the contribution of the critical background, while the second term is the effect of the condensate. Thermodynamically, the condensed phase is therefore characterized
by a negative microcanonical inverse temperature 
\be
\beta_m = \frac{\partial S(a,h)}{\partial (Nh)}\simeq - \frac{1}{\alpha}[(h-h\ss{C})N]^{\frac{1}{\alpha}-1}\,,
\label{eq:beta_m}
\ee
which vanishes in the thermodynamic limit (let us remind that $\alpha>1$).
For $\alpha=2$, the most studied case, the fact that for finite $N$, above $h\ss{C}$ the entropy decreases with increasing the energy can be understood in terms of the effective number $K(h,N)$ of sites hosting the condensate~\cite{GIP21}.
Upon increasing $h$, the extra energy is more and more localized, $K(h,N)$ decreases and the entropy decreases as well, leading to a negative derivative $\partial S/\partial h$. In the thermodynamic limit, however, $K(h,N)=1$ as soon as $h>h\ss{C}$ and adding energy simply adds energy to the single peak, keeping unchanged the entropy. This heuristically justifies the result that $\beta_m=0$ in the whole condensed phase for $N\to\infty$.

A finite $N$ also affects the condensation scenario, as the ``true" localization transition is shifted above the critical line 
by an amount $\delta_N~\sim~N^{(1-\alpha)/(2\alpha-1)}$ for large $N$.\footnote{This result originates from a matching condition between the usual Gaussian scaling of the microcanonical entropy at the critical point $h\ss{C}$, 
$S(a,h)\sim (h-h\ss{C})^2 N $ and the large-deviation scaling of $S(a,h)$ in Eq.~(\ref{eq:LD_entr}), see~\cite{Gradenigo2021_JSTAT}.} In  the  corresponding intermediate range of energies,  $h\ss{C}<h<h\ss{C} +\delta_N$, 
usually referred to as ``pseudo-condensate'' region, spatial localization is practically suppressed by finite-size fluctuations and the
resulting density profiles are effectively delocalized.

This scenario is also visible through the proper order parameter of the transition, the so called energy participation ratio,%
\footnote{In the literature this quantity is also called \textit{inverse} participation ratio.}
\be
Y_2(N) =  \frac{ \left\langle \sum_i \epsilon_i^2\right\rangle}{(hN)^2} ,
\label{eq:Y2}
\ee
where $\langle \cdot \rangle$ represents an ensemble average. For homogeneous states $(h<h\ss{C})$, all sites carry an energy
contribution $\epsilon_i\sim h$, therefore the numerator is extensive and $Y_2$ vanishes in the thermodynamic limit as $ N^{-1}$. 
In the fully localized region, the numerator in~(\ref{eq:Y2}) is dominated by the site $k$ hosting the whole extra energy,
$\epsilon_k= (h-h\ss{C}) N$. As a result, in the thermodynamic limit $Y_2$ converges to a constant: $Y_2= (h-h\ss{C})^2 / h^2 $.   
Finally, in the pseudo-condensate region, the participation ratio is expected to vanish again as $1/N$ due to the delocalized nature 
of this region.
We will come back to this when discussing the nonequilibrium localization.

It is worth stressing that our class of models is of interest both in itself and as a limit of more complicated models. In fact, the Discrete Nonlinear
Schr\"odinger (DNLS) equation~\cite{kevrekidis09} (a Hamiltonian model ubiquitous in physics) displays a non-Gibbsian phase where there is the spontaneous formation of localized excitations. 
This model has two conserved quantities, the total energy $H$ and the total  mass $A=\sum_i c_i$ (also called norm in such a context), where $c_i$ is the square modulus of a complex wave amplitude.
In proximity of the transition line between the homogeneous and the localized phase, local energies are well approximated by a function of local masses,   thereby $H=\sum_i F(c_i)$~\cite{arezzo22}.
In particular, the most studied model, the cubic DNLS equation~\cite{Rasmussen2000_PRL,Iubini2013_NJP,Iubini2017_Entropy,Gradenigo2021_JSTAT,Gradenigo2021_EPJE,arezzo22},
corresponds to $F(c)=c^2$ and it is of interest for different domains, ranging from optics to cold atoms~\cite{kevrekidis09}. However, also the DNLS equations corresponding to $F(c)=c^\alpha$ (nonlinearity of arbitrary order~\cite{johansson04}) and to $F(c)=-\ln (1+c)$ (saturable nonlinearity~\cite{Samuelsen2013}) have been studied. 
The models $F(c)=c^\alpha$ have been studied in~\cite{Szavits2014_PRL} and they are also related to the dynamics of polydisperse hard spheres~\cite{evans10} and to the distribution of entanglement entropy in a random pure state~\cite{Szavits2014_JPA}. 
Finally, at equilibrium these models are also related to the wide class of the Zero Range Processes~\cite{evans05r}, where a single conservation law is present, but a fraction of the mass $c_i$ is transferred from a site $i$ to another site $j$ with a rate depending on $c_i$.

\section{Microscopic dynamics, nonequilibrium setup, heat baths}
\label{sec.baths}

Studying dynamics, either at global equilibrium or in an out-of-equilibrium setup, requires to define some kinetic conservative rules which satisfy detailed balance in the bulk and which are properly modified at boundaries to take into account the possible coupling with heat baths.

The simplest way to satisfy the two conservation rules and the detailed balance condition is to consider a local, stochastic update algorithm: we choose randomly a triplet of consecutive sites $(i-1, i, i+1)$ with local masses $(c_{i-1}$, $c_i$, $c_{i+1})$, and we update it to a new triplet $(c'_{i-1}, c'_i, c'_{i+1})$ such that (i) the sum of the three masses and the sum of the local energies are conserved; (ii) the probability of the transition $\{c\} \to \{ c' \}$ is the same of the inverse transition $\{c' \} \to \{ c\}$. The case where the local energy is the square of the local mass, $F(c)=c^2$, is particularly simple to simulate because the two constraints 
$c_{i-1} + c_i + c_{i+1} =M= c'_{i-1} + c'_i + c'_{i+1}$ 
and
$c^2_{i-1} + c^2_i + c^2_{i+1} = E=
(c'_{i-1})^2 + (c'_i)^2 + (c'_{i+1})^2$
define respectively a plane and a sphere in a three-dimensional space, whose intersection lies on a circle~\cite{Iubini2013_NJP,JSP_DNLS,Szavits2014_JPA,Barre2018_JSTAT}. The accessible mass triplet $(c'_{i-1}, c'_i, c'_{i+1})$ is therefore parameterized by an angle, and the detailed balance condition is easily ensured by picking a random angle extracted from a uniform probability distribution.%
\footnote{For different $F(c)$ the intersection between the two surfaces is no more a circle and the
its parametrization is far from simple, making time consuming the single evolution of the dynamics.}

Depending on the initial masses of the triplet, since $c_i \ge 0$, the physically accessible states can form either a full circle or the union of three disjoint arcs~\cite{JSP_DNLS}. In the latter case, occurring when the mass of a site in the chosen triplet is significantly larger than the other ones, the final state  is chosen in any of the three available arcs. This update allows peaks to freely diffuse in the  lattice~\cite{gotti22}.
For $F(c)\ne c^2$ the solution of the two constraints equations no longer lies on a circle
and a uniform sampling of the microcanonical manifold would require the explicit parametrization of the intersection curve.
In this manuscript we limit numerical simulations to the case $F(c)=c^2$.

Let us now discuss how to couple the system to two heat baths imposing thermodynamic parameters $\mathcal{R}_L=(\beta_L,\mu_L)$ and 
$\mathcal{R}_R=(\beta_R,\mu_R)$ at the left and right chain ends, respectively. If $\mathcal{R}_L=\mathcal{R}_R$, the system attains an equilibrium state, while for $\mathcal{R}_L\neq\mathcal{R}_R$, the system eventually relaxes to an out-of-equilibrium steady state, whose main features will be found analytically in the next Section. 
We will employ models of reservoirs allowing to thermalize the system in any point $(a,h)$ of the homogeneous phase, $h\ss{GS}(a)\le h\le h\ss{C}(a)$.\footnote{This corresponds to $\beta_L,\beta_R\geq 0$ and ensures that the chain edges are always in the homogeneous region. Accordingly, localized states can emerge only in the bulk as a genuine nonequilibrium effect~\cite{gotti22}. } 
To study in detail the dynamics of the nonequilibrium setup, we will use critical heat baths, i.e. reservoirs imposing critical conditions at the chain ends,
$h\ss{C}(a_L)$ and $h\ss{C}(a_R)$, with $a_L\neq a_R$.
This setup allows to obtain a  NESS which is entirely contained in the condensed phase.

Evolution proceeds as follows. We choose at random a site $j = 1,\dots, N$. If an inner site is chosen, $j = 2,\dots, N-1$, then the triplet centered in $j$, $(c_{j-1}, c_j, c_{j+1})$ is updated with the dynamic rule defined above. Conversely, if an outer site $j=1, N$ is chosen, then we only update the local mass $c_j$ of that site to a new one $c_j'$ dependent on the thermodynamic parameters chosen as boundary conditions. 
Boundary updates can be implemented according to a Metropolis grand-canonical rule applied to random perturbations of the initial state $c_j$~\cite{gotti22}. In this paper we consider a more effective strategy that consists in directly imposing  
the equilibrium mass distribution at the chain ends.  In the following we will make explicit reference to the left 
end, where the imposed distribution on the lattice site $c_1$ is
$\rho_L(c_1)=z(\mu_L,\beta_L)^{-1} \exp{[-\beta_L(F(c_1)-\mu_L c_1)]}$. Analogous procedure and definitions  hold for the
site $j=N$ and are readily obtained by replacing $c_1\to c_N$ and $L\to R$.
This method allows for efficient numerical simulations and it is also  useful for the analytical treatment of the nonequilibrium state.

We remark that, for each boundary site, reservoir updates occur on average once every $N$ random updates of the whole lattice, which implies that 
the reservoir efficiency is finite. As a result, usual boundary discontinuities (Kapitza resistance)~\cite{Lepri_PhysRep} manifest themselves 
at the chain ends in steady nonequilibrium conditions. In practice, the actual distributions of $c_1$ and  $c_N$ are found to be slightly different
from $\rho_L(c_1)$ and $\rho_R(c_N)$, respectively. We have verified that this effect has negligible impact on the macroscopic steady state and that it vanishes in the thermodynamic limit.
Boundary discontinuities can be eliminated by adding the 
constraint that every time $c_1$, $c_N$ change with the triplet update they are immediately replaced by new values extracted from the bath distributions. 
We will refer to this boundary rule as heat baths with ``infinite efficiency'', since the ends of the chain are immediately thermalized every time they are moved away from the equilibrium distribution.     
Finally, as usual for Monte Carlo simulations, time is measured in Monte Carlo units, that corresponds to $N$ evolutionary elementary steps.

We can quantify the rate of exchange of mass and energy from the reservoirs to the chain with the appropriate definitions of mass and energy fluxes. For the left boundary, we define\footnote{The quantities appearing in the summations are evaluated every time that site $i=1$ is chosen. On average, this happens every Monte Carlo step.}
\begin{equation}
\begin{split}
   &j_a^{(L)} = \frac{1}{\tau} \sum_{t_k \le \tau} \delta c_1 (t_k)\\
   &j_h^{(L)} = \frac{1}{\tau} \sum_{t_k \le \tau} \delta \e_1 (t_k),
\end{split}
\label{eq:III_fluxL}
\end{equation}
where $ \delta c_1(t_k)$ and $\delta \e_1(t_k) =\delta F(c_1(t_k))$ represent respectively the variations of the mass and energy on the first site  produced by  a reservoir update occurring at time $t_k$. 
Analogous definitions hold for the right boundary and steady transport conditions are attained when $j_a^{(L)}=-j_a^{(R)}$ and   $j_h^{(L)}=-j_h^{(R)}$.

The time $\tau$ appearing in Eqs.~(\ref{eq:III_fluxL}) must formally diverge to obtain steady state currents and the formalism of next Section explains how time averages can be replaced by ensemble averages. If we are interested to the relaxation process towards the steady state, the currents are determined numerically using a large but finite $\tau$.

\section{The Nonequilibrium Steady State: exact spatial profiles and currents}
\label{sec.ness}

In this Section we derive  analytical results for some relevant observables characterizing the NESS, namely average spatial profiles of mass and energies ($a_i$ and $h_i$) and their corresponding parametric profiles ($h_i(a_i)$), mass and energy currents ($j_a$ and $j_h$), and Onsager coefficients.

Let us consider the evolution of the triplet of neighbouring sites centered in $i$, $T_i = (i-1, i, i+1)$, with local masses $(c_{i-1}, c_i, c_{i+1})$. The mass triplet after the update, $T_i'=(c'_{i-1}, c'_i, c'_{i+1})$, is chosen with the constraint that
\bse
\label{eq.triplet}
\begin{equation}
    c'_{i-1} + c'_i + c'_{i+1} = M = c_{i-1} + c_i + c_{i+1}
\end{equation}
\begin{equation}
    \epsilon'_{i-1} + \epsilon'_i+ \epsilon'_{i+1} = E = \epsilon_{i-1} + \epsilon_i+ \epsilon_{i+1},
\end{equation}
\ese
where $\epsilon_i = F(c_i)$. Solutions of Eqs.~(\ref{eq.triplet}) are a one-parameter family of states $T_i'(\theta)=\mathcal{G}(T_i,\theta)$, where
$\mathcal G(\cdot,\theta)$ is a suitable one-to-one transformation mapping $T_i$ to $T_i'$ and $\theta$ is a real parameter.
For $F(c)=c^2$, $\theta$ is an angle~\cite{JSP_DNLS}.

Total variations of mass and energy of site $i$, $\delta c_i=(c_i'-c_i)$ and $\delta \epsilon_i=(\epsilon_i'-\epsilon_i)$, involve the dynamics of three different triplets, namely $T_i$ and $T_{i\pm 1}$.
In addition, two different  kinds of averages must be taken into account: first  the average over $\theta$ for a fixed $T_i$, indicated with the symbol
$\overline{(\cdots)}$, and second the average over the distribution of initial $c_i$, indicated with $\langle(\cdots)\rangle$.

Since Eqs.~(\ref{eq.triplet}) are invariant under any permutation of the final masses or energies of the triplet, the average over $\theta$ is the same for the three sites:
\bse
\label{eq.triplet2}
\be
\overline{c'_{i}}  = \overline{c'_{i-1}} = \overline{c'_{i+1}} = \frac{M}{3},
\ee
\be
\overline{ \epsilon'_{i}} = \overline{ \epsilon'_{i-1}} = \overline{ \epsilon'_{i+1}}= 
\frac{E}{3},
\ee
\ese
therefore giving,  for local masses in triplet $T_i$
\be
\label{eq:dc_bar}
\left( \overline{\delta c_i}\right)_{T_i} =
\tfrac{1}{3}(c_{i-1}+c_i+c_{i+1}) - c_i =
\tfrac{1}{3}(c_{i-1}-2c_i+c_{i+1}).
\ee

If the system is in a steady state, the average over the initial distribution implies that average masses $\langle c_i\rangle$ appearing in the right-hand side
of Eq.~(\ref{eq:dc_bar}) can be replaced by their stationary values $a_i$. Finally, summing over all triplets involving site $i$, we obtain
\begin{equation}
    \langle \overline{\delta c_i }\rangle_{tot} 
    = \frac{1}{3} (a_{i-2} + 2 a_{i-1} -6 a_{i} + 2a_{i+1} + a_{i+2} )
    \quad
    (3\le i \le N-2).
    \label{eq:dc_tot1}
\end{equation}

For sites $i=2$ and $i=N-1$, each of which contained in only two triplets, the total average mass variation is instead given by 
\begin{equation}
\begin{split}
    \langle  \overline{ \delta c_2} \rangle_{tot} &=  \frac{1}{3} (a_{1} -4  a_{2} +  2a_{3} + a_{4} ) \\
   \langle  \overline{ \delta  c_{N-1}} \rangle_{tot} &=  \frac{1}{3} (a_{N} -4  a_{N-1} +  2a_{N-2} + a_{N-3} ).
\end{split}
\label{eq:dc_tot2}
\end{equation}
Clearly, Eqs.~(\ref{eq:dc_tot1}) and (\ref{eq:dc_tot2}) are also valid for the energy profiles, provided that $c_i$ is replaced by $\e_i$ and $a_i$ by $h_i$.

The hypothesis of stationarity implies that total  average variations of mass and energy vanish, 
$\langle \overline{ \delta c_i} \rangle_{tot} = \langle \overline{\delta \epsilon_i} \rangle_{tot} = 0$, therefore from Eqs.~(\ref{eq:dc_tot1}) and (\ref{eq:dc_tot2}) we obtain a system of $N-2$ linear coupled equations for the spatial mass profile,
\bea
    &&a_{1} -4  a_{2} +  2a_{3} + a_{4} = 0\nonumber\\
    &&a_{i-2} + 2 a_{i-1} -6 a_{i} + 2a_{i+1} + a_{i+2}  = 0
    \quad (3\le i\le N-2)\nonumber\\
    &&a_{N} -4  a_{N-1} +  2a_{N-2} + a_{N-3} = 0.
\eea
An identical (and independent) system of equations is valid for the energies $h_i$.

The missing two equations depend on the coupling with heat baths, see Sec.~\ref{sec.baths}. If they have ``infinite efficiency'' (IE), mass boundary conditions are 
\be
a_1 = a_L, \qquad a_N = a_R \quad \mbox{(IE baths)}\,,
\ee
where the mass parameters $a_L$ and $a_R$ are completely defined through equilibrium grand-canonical relations respectively by $(\mu_L,\beta_L)$ and $(\mu_R,\beta_R)$, see Eq.~(\ref{eq:state}) in Appendix~\ref{app.mc}. 
In the presence of  heat baths with ``finite efficiency'' (FE), additional equations obtained from the average mass variation on sites $i=1,N$ must be considered, explicitly $\langle \overline{\delta c_{1,N} }\rangle_{tot} = \langle \overline{\delta c_{1,N}} \rangle_{T_{2,N-1}} + (a_{L,R} - a_{1,N}) = 0$, from which we get  
\begin{equation}
\begin{split}
    & 3 a_L - 5a_1 + a_2 + a_3 = 0 \\
    & 3 a_R - 5a_N + a_{N-1} + a_{N-2} = 0
\end{split}
\quad\mbox{(FE baths)}.
\end{equation}
Again, analogous equations for the energy hold replacing $a_{L,R}$ by $h_{L,R}$ and $a_i$ by $h_i$.

Once the boundary conditions have been chosen, average spatial profiles of mass $a_i$ and energy $h_i$ of the stationary state can be obtained by solving the two systems of equations, whose solutions can be expressed in the following form
\bse
\label{eq.aihi}
\begin{equation}
    a_i = a_L + A_i (a_R - a_L)
\end{equation}
\begin{equation}
    h_i = h_L + A_i (h_R - h_L).
\end{equation}
\ese
The real coefficients $A_i$, which take values between 0 and 1, are the same for mass and energy because $a_{L,R}$ and $h_{L,R}$ do not explicitly appear in the equations: coefficients $A_i$ depend only on the position $i$ and the size $N$ of the system. Furthermore,  since the systems of equations are invariant under the exchange $L\to R$ and $i\to N+1-i$, we have that $A_{i} = 1 - A_{N+1-i}$. Explicit expressions  for $A_i$ are derived in Appendix~\ref{app:Ai}.

    \begin{figure}
        \includegraphics[width=1.0\columnwidth]{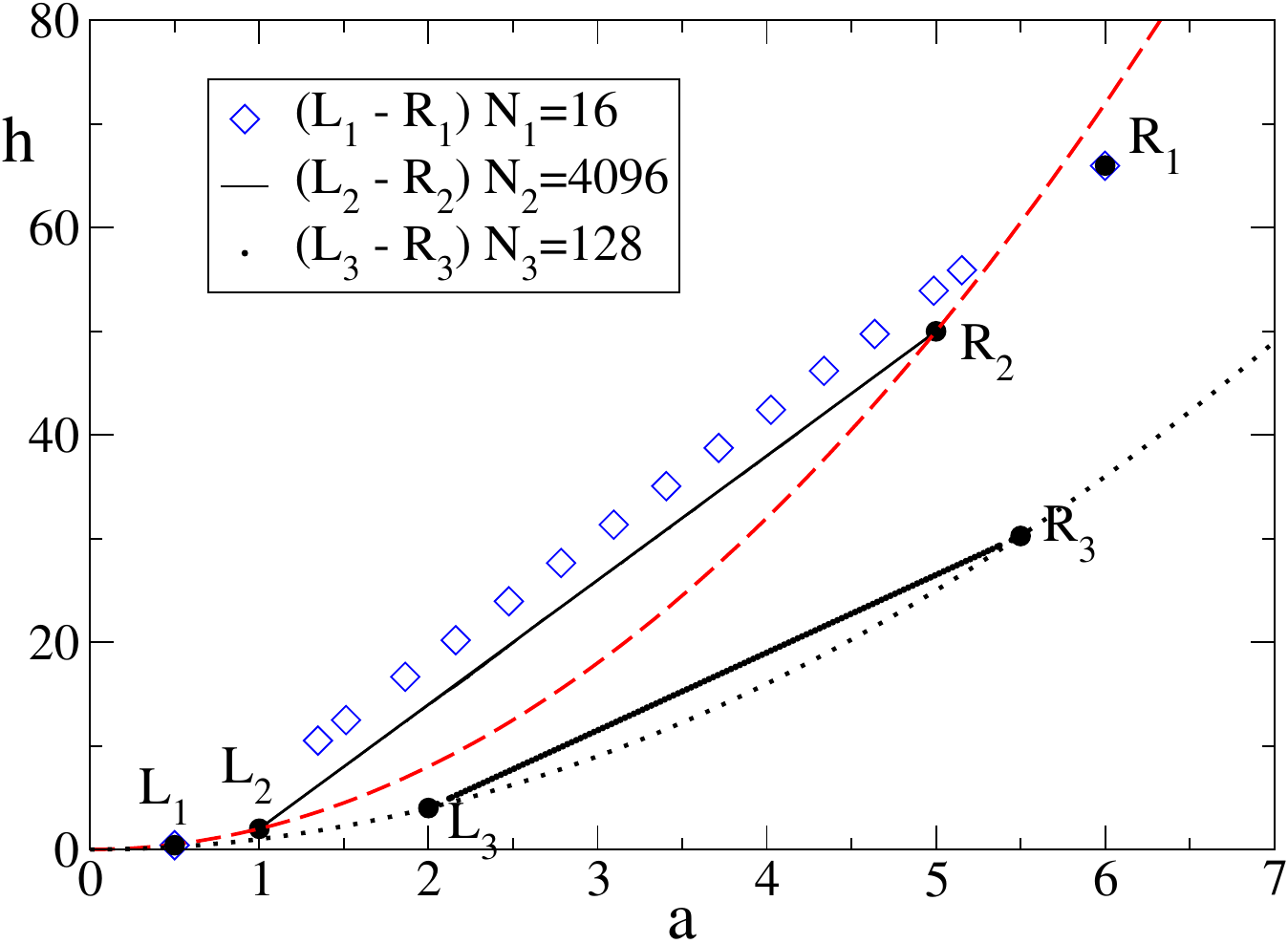} 
        \caption{Parametric profiles with different boundary conditions $(L_i - R_i)$ and different system sizes $N_i$, see the legend. The dotted line and the dashed red line are, respectively, the ground state line and the critical line, for $F(c)=c^2$.   
        The first setup ($i=1)$ corresponds to IE baths, both located in the homogeneous region ($a_L=0.5$, $h_L=0.41$, $a_R=6$, $h_R=66$). The second setup ($i=2$) corresponds to FE critical baths. The third setup ($i=3$) corresponds to FE baths, both located on the ground-state curve.
        Data have been obtained numerically and they coincide with the expected analytical expressions.
        }\label{fig:ah_profile_np}
    \end{figure}

A first implication of Eqs.~(\ref{eq.aihi}) is that parametric profiles are perfectly linear independently on the boundary parameters and on the system size. In fact,  we obtain that points $(a_i,h_i)$ are arranged along the straight line
\be
\label{eq:lin_ha}
h(a) = h_L + \frac{h_R-h_L}{a_R-a_L} (a-a_L) ,
\ee
as shown in Fig.~\ref{fig:ah_profile_np}, where we plot the parametric curves for three different nonequilibrium setups. We remark that the spatial and parametric profiles do not depend on the choice of $F(c)$, but only on the location of the baths parameters in the $(a,h)$ plane, on the finite/infinite efficiency of the baths themselves, and on the size $N$.  

\begin{figure}
        \includegraphics[width=1.0\columnwidth]{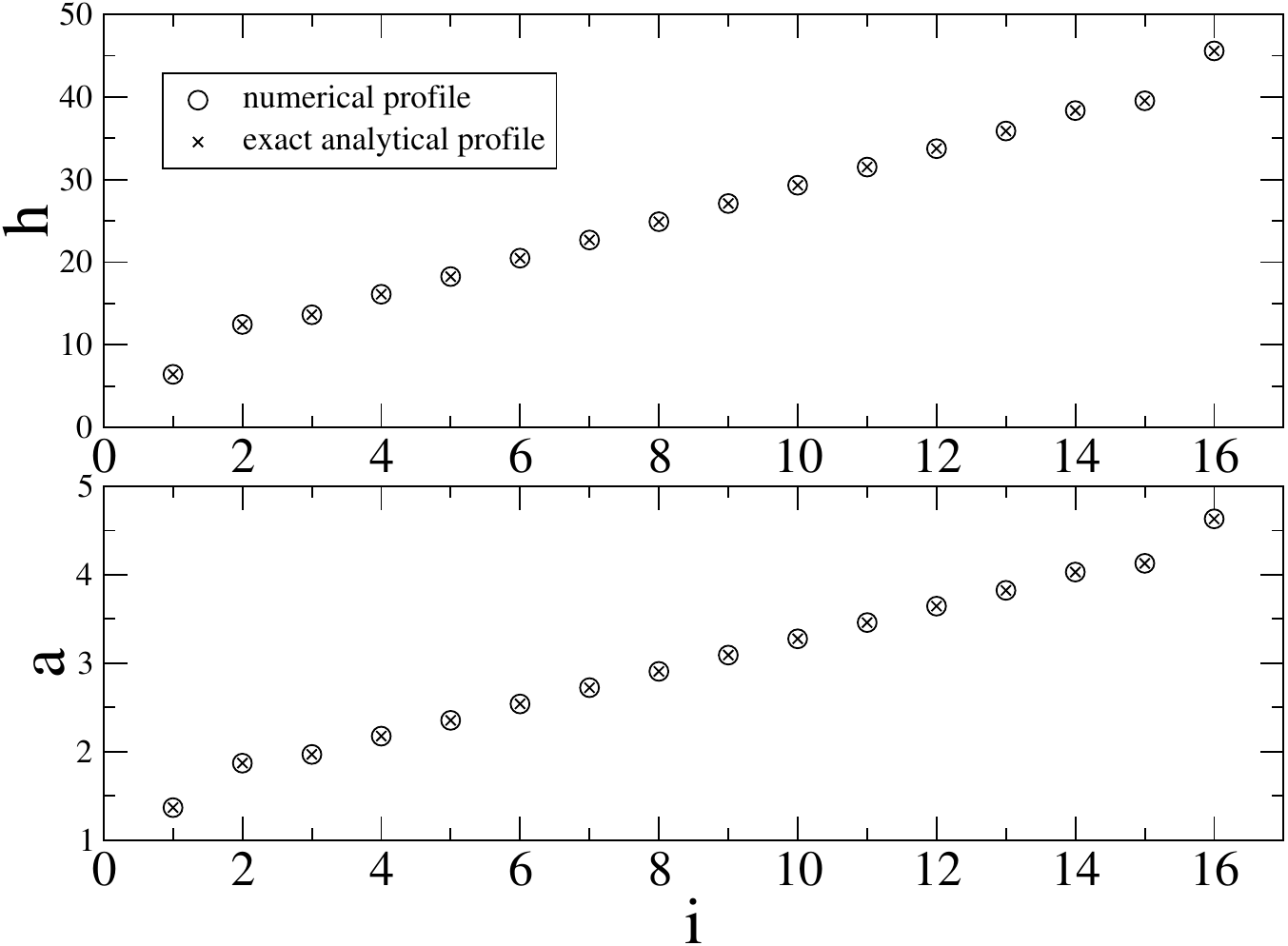} 
        \caption{Model $F(c)=c^2$. Analytical and numerical spatial profiles of mass (bottom panel) and  energy (top panel), for $N = 16$ and boundary conditions on the critical line $h\ss{C} = 2a^2$. ``Finite efficiency'' (FE) reservoirs are employed with  $a_L = 1$ and $a_R = 5$. 
        }\label{fig:profilispaziali1}
    \end{figure}

While parametric profiles are linear for any $N$, spatial profiles are affected by boundary resistance effects, so that $a_i$ and $h_i$ are linear only for large $N$.
However, as shown in Fig.~\ref{fig:profilispaziali1} it is apparent that even for a value of $N$ as small as 16, heat baths modify very weakly the linear profile. 

Let us now determine the mass and energy currents in the NESS. 
Using Eqs.~(\ref{eq:III_fluxL}), (\ref{eq.aihi}), and (\ref{eq:app_A3}), currents for FE baths read 
\begin{equation}
    j_a^{(L)} = \langle\delta c_1\rangle = a_L - a_1 = -A_1 (a_R - a_L)
\end{equation}
\begin{equation}
\label{eq:jhFE}
    j_h^{(L)} = \langle\delta \e_1\rangle = h_L - h_1 = - A_1 (h_R - h_L)
\end{equation}
\begin{equation}
    j_a^{(R)} =\langle\delta c_N\rangle = a_R - a_N = A_1 (a_R - a_L) = - j_a^{(L)}  
\end{equation}
\begin{equation}
    j_h^{(R)} = \langle\delta \e_N\rangle =h_R - h_N =  A_1 (h_R - h_L) = - j_h^{(L)} .
\end{equation}
Similar calculations for IE reservoirs give
\begin{equation}
    j_a^{(L)} = - \frac{1}{3}(A_2 + A_3) (a_R - a_L)
\end{equation}
\begin{equation}
    j_h^{(L)} = - \frac{1}{3}(A_2 + A_3) (h_R - h_L).
\end{equation}

In the limit of large $N$, for FE baths, $A_1 \sim 2/N$, and for IE baths, 
$(A_2 + A_3)/3 \sim  2/N$, so that $j_{a,h}$ do not depend on the efficiency of the reservoirs and we obtain
\begin{equation}
    j_a =  j_a^{(L)} = -j_a^{(R)} \simeq -\frac{2}{N} (a_R - a_L) \simeq -2 \partial_x a  
    \label{eq.ja}
\end{equation}
\begin{equation}
    j_h = j_h^{(L)} = -j_h^{(R)} \simeq -\frac{2}{N} (h_R - h_L) \simeq -2 \partial_x h  .
    \label{eq.jh}
\end{equation}
The $1/N$ dependence of stationary currents on boundary imbalances of mass and energy in Eqs.~(\ref{eq.ja}) and (\ref{eq.jh}) shows that transport is always diffusive, irrespective of possible condensation processes occurring in the lattice. This result is less straightforward than one can imagine~\cite{Iubini2017_Entropy}.

\begin{figure}
        \includegraphics[width=1.0\columnwidth]{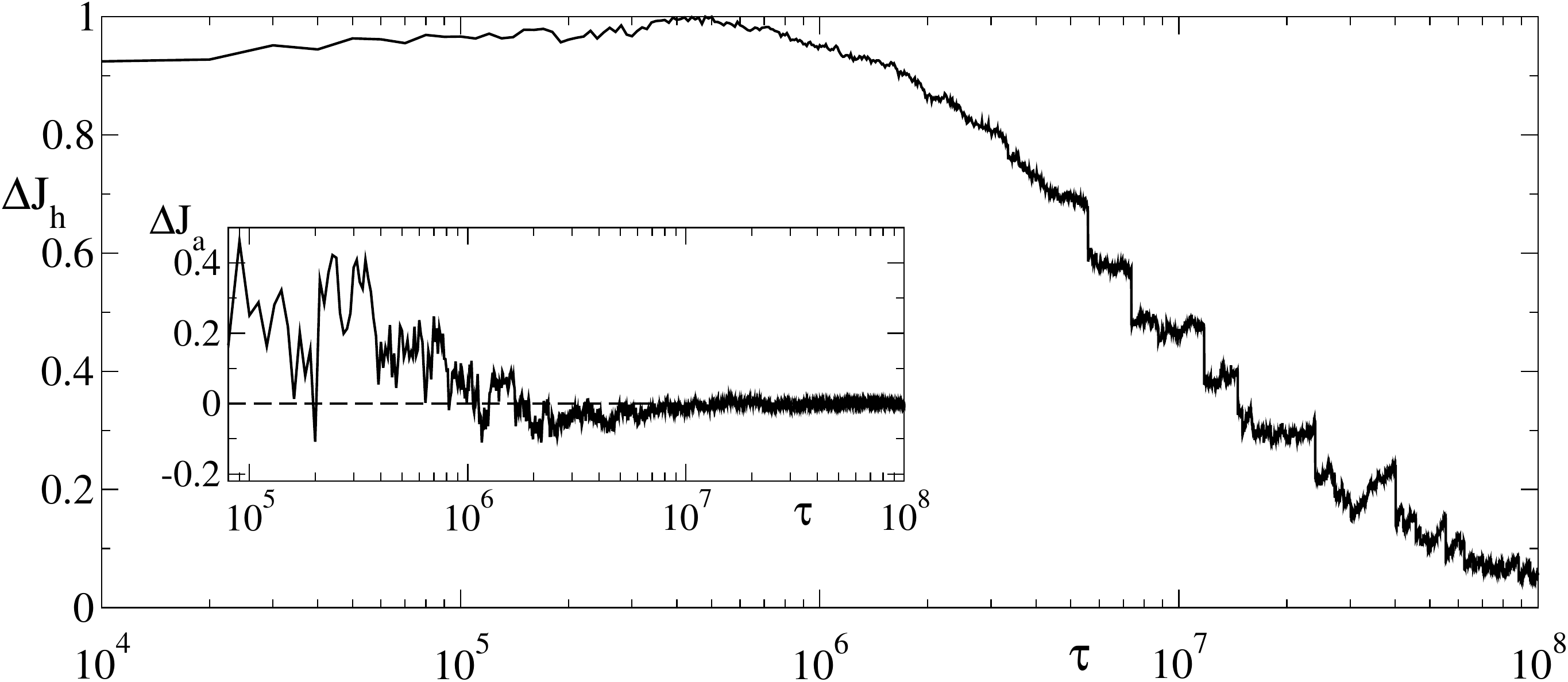} 
        \caption{Model $F(c)=c^2$. Evolution of the relative asymmetry of the energy flux $\Delta J_h (\tau)$ for a system of size $N=8192$. The system is in contact with FE reservoirs imposing  boundary conditions on the critical line, with  $a_L = 1, a_R = 5$. Inset: evolution of the relative asymmetry of the mass flux, $\Delta J_a(\tau)$ .}
        \label{fig:relat_asym}
    \end{figure}

Numerical simulations confirm the above predictions. As a first test, we  verified the convergence to a NESS for a system driven in the localized region of parameters. 
In Fig.~\ref{fig:relat_asym} we plot the evolution of the relative asymmetry of the left/right currents of mass and energy, namely
\be
\Delta J_{a,h} (\tau) = \frac{(|j_{a,h}^{(R)}(\tau)| - |j_{a,h}^{(L)}(\tau)|)}{(|j_{a,h}^{(R)}(\tau)| + |j_{a,h}^{(L)}(\tau)|)}\,,
\ee
which must vanish in a NESS. 
The figure shows that the asymmetry  of the energy current $\Delta J_{h}$ (main panel) vanishes much more
slowly than the corresponding mass asymmetry  $\Delta J_{a}$ (inset). 
This is due to the fact that when the parametric curve overpasses the critical line and enters the condensed region, emerging peaks localize energy, not mass.\footnote{This is true for a positive $F(c)$, see the last Section for a comment on this when $F(c)$ is negative.}
A more detailed discussion of the localization properties in the NESS is contained in Sec.~\ref{sec.differences}.

\begin{figure}
        \includegraphics[width=1.0\columnwidth]{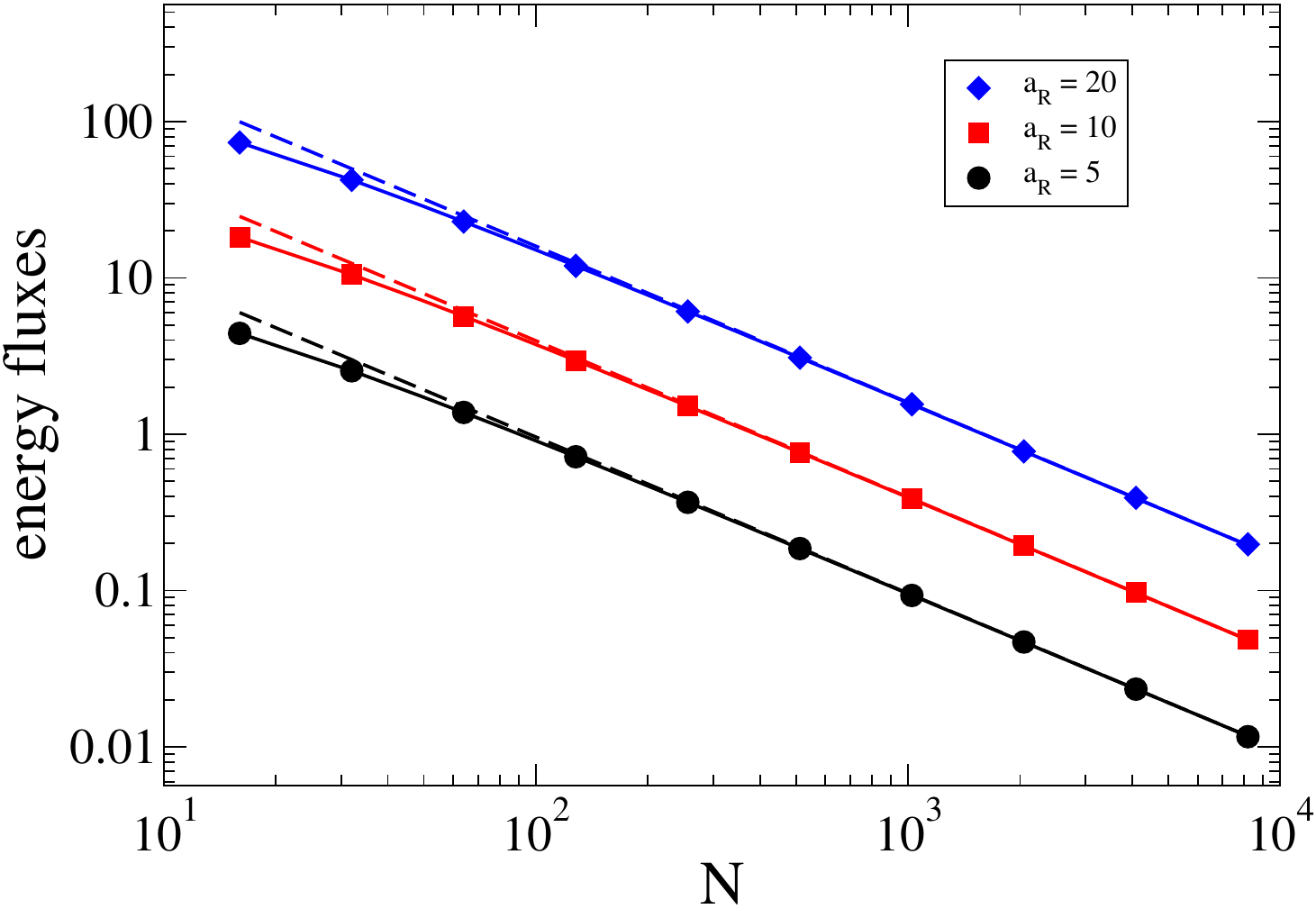} 
        \caption{Stationary energy currents
        versus $N$. Boundary conditions are obtained  with FE critical heat baths imposing parameters $ a_L = 1$ and different values of $a_R$, see legend. Full symbols indicate the values of the fluxes obtained numerically for a total simulation time $\tau = 6.6 \times 10^9$.
        Solid lines indicate the exact analytical analytical expression, see Eq.~(\ref{eq:jhFE}).
        Dashed lines are the asymptotic expressions $|j_h| = \frac{2}{N} (h_R - h_L)$, see Eq.~(\ref{eq.jh}). 
        }\label{fig:Flussoenergia}
    \end{figure}

In Fig.~\ref{fig:Flussoenergia} we show the behavior of  energy currents  for different system sizes $N$ in the long-time limit $\tau\gg 1$.
The good agreement between numerics (symbols) and the stationary analytical expression (full lines) attests that a NESS has always been attained in simulations, while the comparison with the asymptotic expressions for  large $N$ (dashed lines) shows that as for the currents the thermodynamic limit is attained already for relatively small sizes.

Finally, the knowledge of explicit expressions relating stationary currents to macroscopic mass and energy gradients, see Eqs.~(\ref{eq.ja}) and (\ref{eq.jh}),
allows us to derive complete information on transport coefficients in the linear-response regime, which can be expressed in terms of Onsager coefficients~\cite{livi17}. 
More precisely, we can rewrite Eqs.~(\ref{eq.ja},\ref{eq.jh}) in the form
\begin{equation}
\begin{split}
    &j_a =  C_{aa} \partial_x a + C_{ah} \partial_x h\\
    &j_h =  C_{ha} \partial_x a + C_{hh} \partial_x h,\\
\end{split}
\end{equation}
with constant coefficients
\begin{equation}
    \begin{split}
        &C_{aa} = C_{hh} = -2 \\
        &C_{ah} = C_{ha} = 0 \,.
    \end{split}
\end{equation}
We  recall that Onsager coefficients are properly defined in terms of gradients of $\beta$ and $m=\beta\mu$~\cite{iubini23} 
\begin{eqnarray}
\label{eq.Ons1}
j_a &=& -L_{aa}\, \p_x m + L_{ah}\p_x\beta \\
j_h &=& -L_{ha}\, \p_x m + L_{hh}\p_x\beta\,.
\label{eq.Ons2}
\end{eqnarray}
Therefore, passing from $(a,h)$ to $(m,\beta)$ we obtain
\begin{eqnarray}
\label{eq:Lxy}
L_{aa}=2\partial_m a \quad  &L_{ah}&=-2\partial_\beta a,\\\nonumber
L_{ha}=2\partial_m h \quad  &L_{hh}&=-2\partial_\beta h ,
\end{eqnarray}
where the derivatives appearing in $L_{ij}$ can be derived
from equilibrium relations $a=a(m,\beta)$ and $h=h(m,\beta)$, see Eq.~(\ref{eq:state}). As expected,  the Onsager coefficients in Eq.~(\ref{eq:Lxy}) 
satisfy the celebrated Onsager reciprocity relations, $L_{ah}=L_{ha}$. Indeed, recalling the grand-canonical relations $a=\partial_m \log{z}$
and $h=-\partial_\beta \log{z}$, see Appendix~\ref{app.mc}, the equality of off-diagonal coefficients follows from 
$\partial_m\partial_\beta\log{z}=\partial_\beta\partial_m\log z$ and from the regularity of the partition function $z(m,\beta)$ in the homogeneous region.
In this respect, it is worth noticing that the above derivation of transport coefficients is necessarily restricted to the homogeneous region $\beta\geq 0$, as it
requires the existence of the grand-canonical ensemble. 
Although the problem of energy transport at negative absolute temperatures was recently explored in~\cite{baldovin21,Baldovin_2021_review}, we do not
expect  straightforward generalizations for $\beta<0$ for the class of condensation models here studied. The main reason
is that such a program would require to connect the system to negative-temperature reservoirs, thereby typically resulting in condensation instabilities and absence of stationary conditions. 

It is interesting to work out the limit of vanishing $\beta$ of $L_{ij}$ in Eq.~(\ref{eq:Lxy}) for the class of models $F(c)=c^\alpha$.  In this case the critical curve is $h\ss{C}(a)=\Gamma(1+\alpha)a^\alpha$ and we obtain (see Appendix~\ref{app.Onsager})
\begin{eqnarray}
\label{eq.Lij_alpha}
L_{aa}= 2a^2 \quad  &L_{ah}&= 2\alpha \Gamma(\alpha +1) a^{\alpha +1},\nonumber \\
L_{ha}= 2\alpha \Gamma(\alpha +1) a^{\alpha +1} \quad  &L_{hh}&= 2\left[\Gamma(2 \alpha + 1) - \Gamma^2(\alpha +1)\right]a^{2\alpha }\,.
\end{eqnarray}
Therefore, all Onsager coefficients are well defined and positive even on the critical line. Moreover, from the definition of the Seebeck coefficient~\cite{benenti17rev}  
\begin{equation}
    \mathcal{S} \equiv \beta \frac{L_{ah}}{L_{aa}} - m\,,
\end{equation}
we obtain that on the critical line, $\mathcal{S}=-m=1/a$. Since $\mathcal{S}\neq 0$, we can conclude that mass and energy currents in $\alpha$-models are coupled in the usual sense of  irreversible thermodynamics.  These results generalize the study of Ref.~\cite{iubini23} on Onsager coefficients, which was restricted to the case $\alpha=2$.

\section{Statistical and dynamical properties of nonequilibrium condensation}
\label{sec.differences}

In this section we discuss in more detail the out-of-equilibrium condensation process and we compare it with the analog phenomenon occurring at equilibrium. For simplicity we refer to the class $F(c)=c^\alpha$.

Equilibrium condensation can only appear in the microcanonical statistical ensemble, because the region above the critical line, $h>h\ss{C}(a)$, 
 cannot be described in terms of the grand canonical ensemble. The equilibrium condensed phase is therefore characterized by the exact conservation of the energy density $h$, which can be decomposed as the sum of the critical energy density $h\ss{C}$ plus an extra energy density $\Delta=h-h\ss{C}$.
The former is uniformly distributed in the whole system, according to an exponential distribution of the mass; the latter is localized on a not-extensive number of sites, $K(N,\Delta)$~\cite{GIP21}. In the thermodynamic limit,
$K(N\to \infty,\Delta)\to 1$ independently on $\Delta$: only one site hosts the entire extra-energy, giving rise to a Dirac-delta peak in the energy distribution $p(\epsilon)$, for $\epsilon=\Delta N$: 
$p(\e) = p_c(\e) + \tfrac{1}{N}\delta(\e -\Delta N)$.
The critical energy density $p_c(\e)$ is explicitly found from the relation $p_c(\e) d\e = \rho_c(c) dc$, where $\e=F(c)$ and $\rho_c(c)=(1/a)e^{-c/a}$. For $F(c)=c^\alpha$, 
\be
p_c(\e) = \frac{\exp{\big(-\e^{1/\alpha}/a\big)}}{a\alpha \e^{1-1/\alpha}} .
\label{eq.pce}
\ee

\begin{figure}
        \includegraphics[width=1.0\columnwidth]{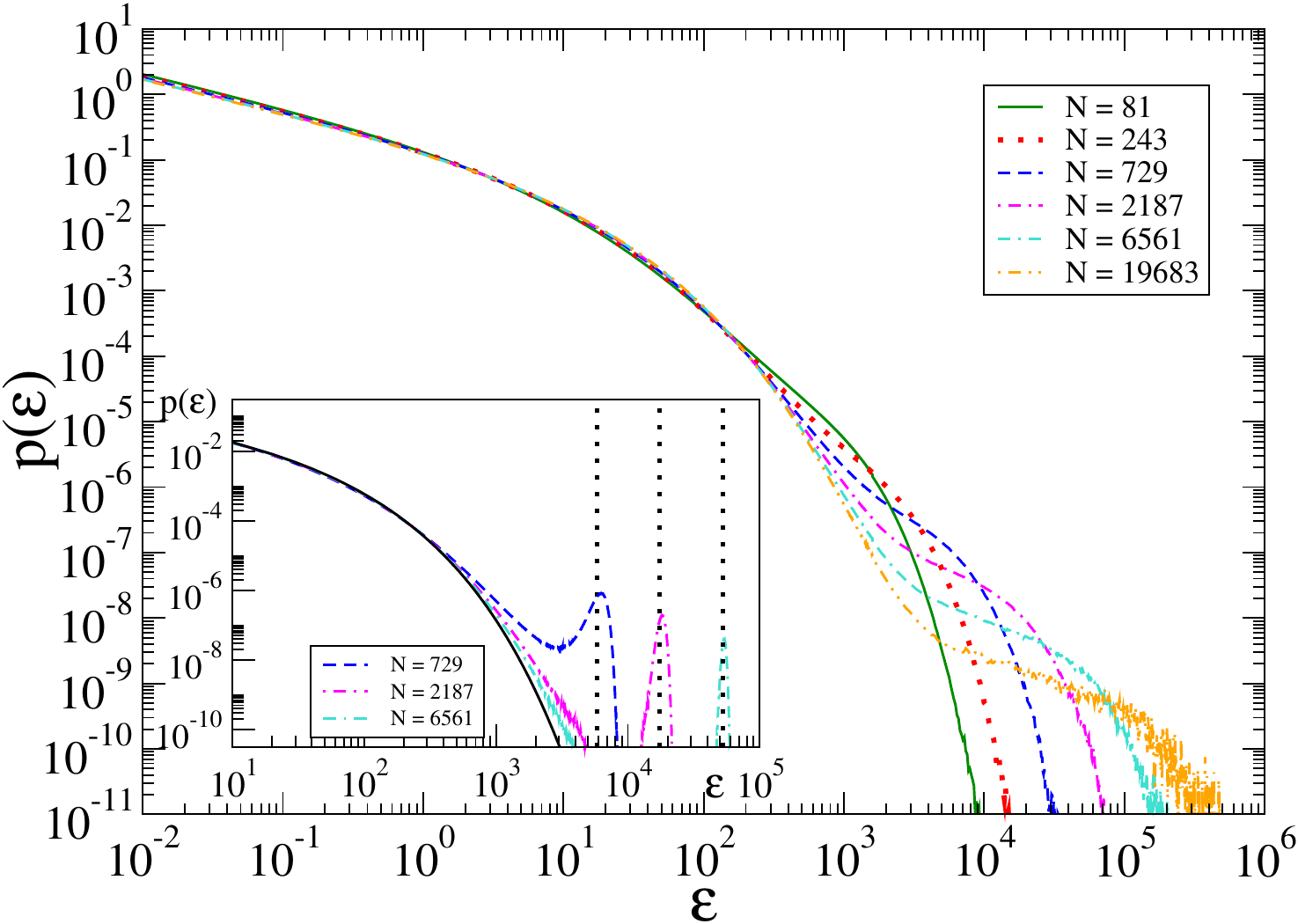} 
        \caption{Main: Nonequilibrium setup with critical baths imposing $a_L=1$ and $a_R=5$. We plot the stationary energy distribution $p(\e)$ for the central site of systems with increasing size $N$. The size is chosen so that its average mass is $a=3$ and its average energy is $h=26$. The NESS state has been sampled for times  much larger than the typical peak's lifetime $t_p(N)$. For comparison, in the inset we plot the equilibrium energy distribution for the same values of $a$ and $h$. The full black line is the critical distribution Eq.~(\ref{eq.pce}) for $\alpha=2$ and $a=3$. The vertical dotted lines locate the expected energy of the condensate, equal to $(h-h\ss{C}(a))N$.}
        \label{fig:cd_energy}
    \end{figure}

For finite $N$, the condensate contribution to $p_c(\e)$ broadens to a finite-height bump, see the inset of Fig.~\ref{fig:cd_energy}, where we show the numerical equilibrium energy distributions in the condensed region for $\alpha=2$ and different sizes $N$. 
More generally,  finite-size effects were found to deeply modify the equilibrium localization scenario expected in the thermodynamic limit~\cite{Szavits2014_PRL,Gradenigo2021_JSTAT,GIP21}. Among the most important effects, there is the emergence of a ``pseudo-localized'' region above the critical line $h\ss{C}(a)$ in which the system is effectively delocalized.

    \begin{figure}
        \includegraphics[width=1.0\columnwidth]{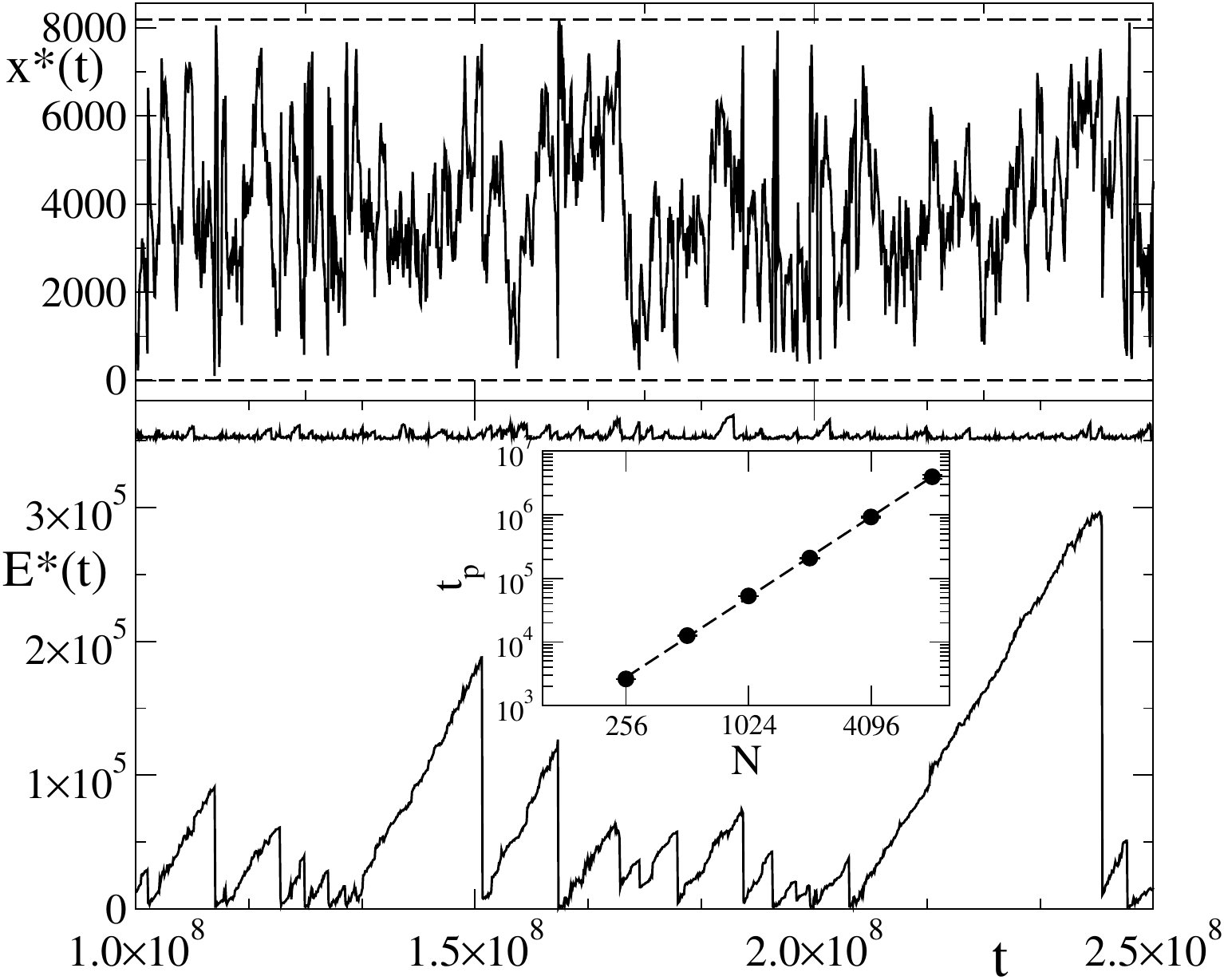} 
        \caption{Top panel: Evolution of the main peak's position $x^*(t)$, after a transient $\tau_0=10^8$, on a system with size $N = 8192$. The two horizontal dashed lines indicate the borders of the system, which interacts with two critical heat baths imposing $a_L = 1$ and $a_R = 5$.
        Bottom panel: the energy $E^*(t)$ of the main peak (lower curve) and the energy of the second highest peak, vertically translated to make it evident.
        In the inset we plot the mean lifetime $t_p$ of peaks as a function of the system size $N$. The power-law fit, $t_p \simeq 0.025 N^{2.01}$, reveals the diffusive character of their motion.}\label{fig:en_pos_8k}
    \end{figure}

Coming now to the out-of-equilibrium setup with critical heat baths, we can argue that dynamical evolution proceeds repeatedly through the onset 
of travelling peaks that are eventually destroyed at the chain ends, as their presence is not compatible with heat baths at positive or infinite temperature.
This is clearly shown in Fig.~\ref{fig:en_pos_8k}, where we plot the time dependence of position (upper panel) and energy (lower panel) of the highest peak detected in the chain at various times during a long evolution. The steady state thereby manifests itself as a balance of the creation and death processes of such peaks. Therefore, if $t_p(N)$ is the typical lifetime of a peak, a NESS establishes only on timescales $t\gg t_p(N)$. On timescales shorter or of the order of $t_p(N)$ the system is not stationary: there is an unbalance of left and right currents which induces the growth of the travelling peaks, which appears to be linear in time.
From Fig.~\ref{fig:en_pos_8k} we have also clear numerical evidence that $t_p(N)$ increases as $N^2$, see the inset of the bottom panel: this result signals a symmetric spatial diffusion process of peaks.
In fact, we have not found evidence of drifts towards a specific side of the chain. Nevertheless, we expect that such unbiased diffusion is not universal, as it eventually depends on the underlying microscopic dynamical rule: an example showing biased diffusion of peaks in out-of-equilibrium conditions was found in~\cite{Iubini2017_Entropy} for the deterministic DNLS equation.

At first glance, the emerging  picture of an out-of-equilibrium condensed phase originated by traveling peaks superposed to a delocalized background might appears comparable to the equilibrium one. On the other hand, a closer inspection reveals an important difference, because in the 
NESS peaks are repeatedly created in the bulk and destroyed at the system's boundaries and their energy increases in time until they disappear, while in equilibrium conditions their energy simply fluctuates around an average value.
Therefore, in nonequilibrium conditions a single site experiences the passage of peaks of variable heights, even if the number of peaks which are present in the system at a given time is of order one. 
The above picture of peaks diffusing and carrying  extra energy with respect to the average local critical energy can be quantitatively supported, firstly, by a numerical evaluation of the average energy transported by peaks; secondly, by an alternative derivation, within a continuum approximation, of the parametric curve of the NESS.

The global extra energy is the sum $E_{ex} =\sum_i (h_i -h\ss{C}(i))$, where $h_i$ is the average energy of site $i$ and $h\ss{C}(i) =h\ss{C}(a_i)$ is the critical energy corresponding to the average mass of site $i$. We claim that $E_{ex}$ corresponds to the time average of the energy of traveling peaks. As a numerical test, we can refer to the setup of Fig.~\ref{fig:en_pos_8k} and limit ourselves to the two highest peaks shown in the bottom panel. Remembering that for large $N$, $a_i$ and $h_i$ are linear profiles and that $h\ss{c}(i)=2a_i^2$, we obtain 
\be
\label{eq:Eex}
E_{ex} \simeq \int_0^N dx\, 2(a_R -a(x))(a(x) -a_L) \simeq 4.37\times 10^5,
\ee
where $a(x)=a_L +\tfrac{x}{N}(a_R-a_L)$, $a_L=1$, $a_R=5$, and $N=8192$. 
On the other hand, the time-average energies of the main and second peak are respectively  $3.94\times 10^5$ and $0.38\times 10^5$.
Their sum is $4.32\times 10^5$, in fairly good agreement with the value of $E_{ex}$ found in Eq.~(\ref{eq:Eex}).

Passing to a continuum description, it is also possible to evaluate the spatial density of the extra energy, $\Delta(x)=h(x)-h\ss{C}(x)$, which is supposed to satisfy a stationary, diffusion equation of the following form,
\be
D \frac{d^2 \Delta(x)}{dx^2} + I(x) = 0,
\label{eq.diffusion}
\ee 
where $D=2$ is the diffusion coefficient of peaks, derived in Appendix~\ref{app.D}, and $I(x)$ is a source term accounting for energy injection, determined assuming a ``critical" background. If $a(x)$ is the average mass density at site $x$, the energy background is $h\ss{C}(a(x))$, and $I(x)$ is determined through the total variation of energy at the site $i$ located in $x$, once we average over all moves involving site $i$,\footnote{See Eq.~(\ref{eq:dc_tot1}), replacing $c_i$ with $\e_i$ and $a_i$ with $h_i$.}
\begin{equation}
    \langle \overline{\delta \e_i} \rangle_{tot}
     = \frac{1}{3} (h_{i-2} + 2 h_{i-1} -6 h_{i} + 2h_{i+1} + h_{i+2} ).
\end{equation}
In the steady state the left hand side vanishes, which imposes the equation determining the spatial profile $h_i$. Here instead we suppose that $h_i$ is the local, critical energy, so that the right-hand side does not vanish and the left-hand side is exactly the source term $I(x)$.
Passing from discrete to continuous, we obtain
$I(x) = 2 d^2 h\ss{C}(x)/dx^2$, and Eq.~(\ref{eq.diffusion}) reads
\be
\label{eq.diffusion2}
2 \frac{d^2}{dx^2} \left[ \Delta(x) + h\ss{C}(x) \right] = 0,
\ee
which must be solved with  absorbing boundary conditions $\Delta(0)=\Delta(L)=0$.
We therefore obtain that $\Delta(x)+h\ss{C}(x)$ has a linear profile passing from the values imposed by critical baths, in agreement with the
exact, discrete microscopic description discussed in Sec.~\ref{sec.ness}.

    \begin{figure}
        \includegraphics[width=1.0\columnwidth]{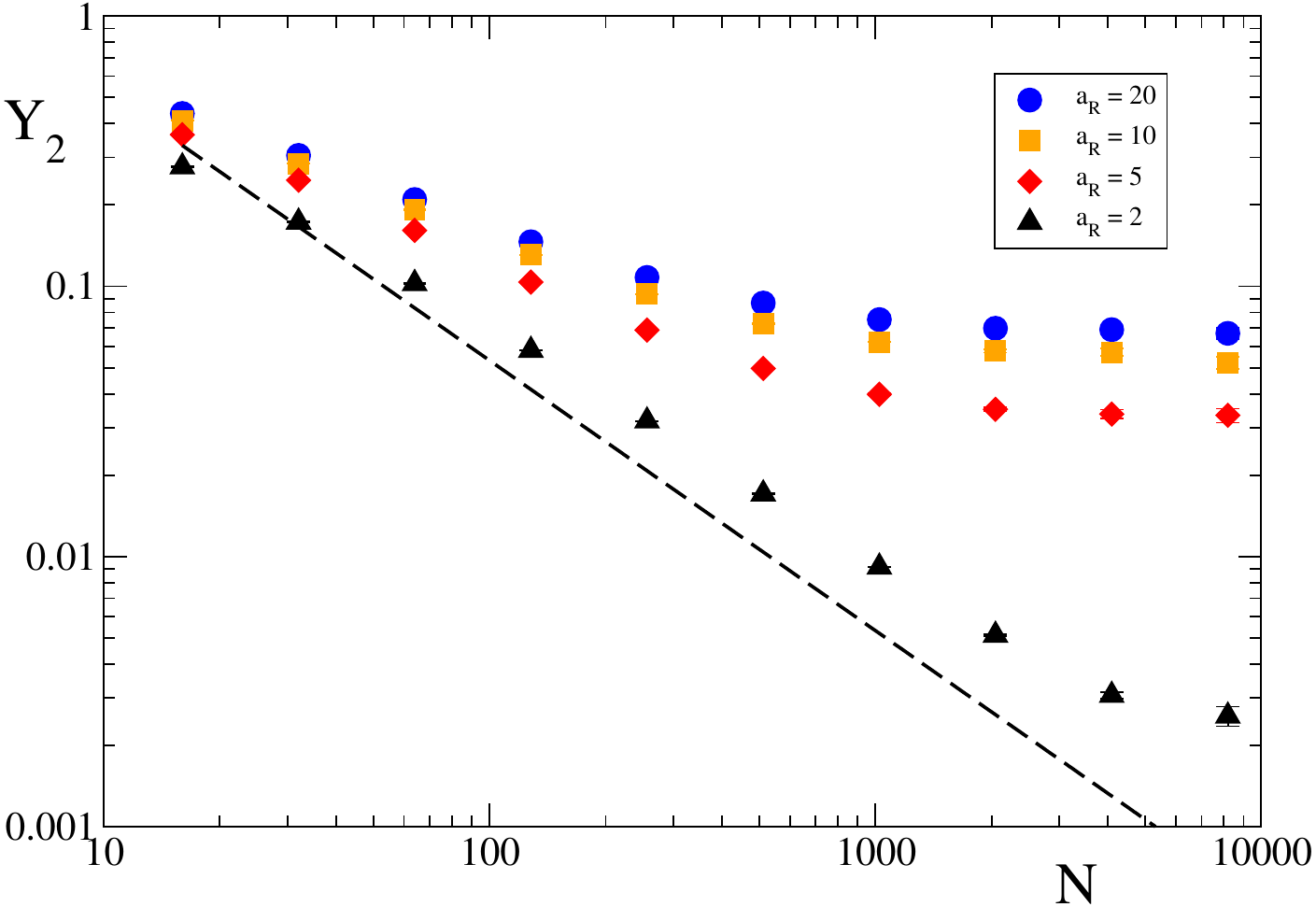} 
        \caption{Participation ratio $Y_2$  vs system size $N$ for a NESS obtained with critical heat baths imposing $a_L = 1$ and different values of $a_R$ (see legend). $Y_2$ is obtained numerically from a temporal average over a time $\tau = 6.6 \times 10^9$. The uncertainty on $Y_2$ is computed as the standard deviation $\sigma = \sqrt{\overline{Y_2^2}  - \overline{Y_2}^2 }$, divided by $\sqrt{\tau/\tau_{ac}}$, where $\tau_{ac}$ is the autocorrelation time of the signal $Y_2(t)$. The dashed line is a reference power-law decay $\sim1/N$.}  \label{fig:Y2}
    \end{figure}

The different dynamical behavior of equilibrium and out-of-equilibrium condensation manifests itself also in the stationary energy distribution, $p(\e)$ on a single site. A comparison of the two regimes is shown in Fig.~\ref{fig:cd_energy}, where we plot the equilibrium (inset) and out-of-equilibrium (main panel) energy distributions corresponding to the same values of mass and energy densities.
The equilibrium distribution has the expected $N-$dependent form: for $\e \ll N\Delta$, $p(\e) \simeq p_c(\e)$, with $p_c(\e)$ given in Eq.~(\ref{eq.pce}), while, for larger energies,  $p(\e)$ displays a bump around $\e=N\Delta$. The width of the bump progressively decreases 
upon increasing $N$.
On the other hand, the nonequilibrium distribution does not display any bump: it is a monotonically decreasing function, which however extends over increasingly larger energies for increasing $N$.

The lack of the condensation bump in the NESS energy distribution affects also the behavior of the participation ratio $Y_2(N)$ defined in Eq.~(\ref{eq:Y2}), which is the usual order parameter for the condensation transition.
In equilibrium conditions and for finite sizes, $Y_2(N)$ displays a minimum when plotted versus $N$, for sufficiently small values of  $\Delta =h-h_c$. 
Such a minimum originates essentially from the fact  that the condensate is localized on a small number of sites and that the extra energy is constant in time, apart from  statistical fluctuations~\cite{GIP21}. The regime of nonequilibirum condensation is shown in Fig.~\ref{fig:Y2},
where we plot $Y_2(N)$ for different nonequilibrium critical boundary conditions with fixed $a_L=1$ and different $a_R$.
We always find that $Y_2$ decreases monotonically with $N$ and we attribute this feature to the peculiar dynamics of growth and absorption of travelling peaks.
Although we explored the phenomenon for relatively large sizes and times, numerics by itself is not conclusive and further efforts would be
in order to clarify asymptotic behaviors, in particular to understand how $Y_2(N\to\infty)$ depends on $a_{L,R}$ when $(a_R - a_L)\to 0$, the nonequilibrium equivalent of $\Delta\to 0$ for equilibrium systems.

\section{A critical summary}
\label{sec.discussion}

We have analyzed the phenomenon of nonequilibrium condensation in a class of open lattice models with two conservation laws steadily driven by boundary reservoirs. Qualitatively new features were found with respect to the usual condensation transition occurring in equilibrium
conditions. The most important one is that nonequilibrium condensation can appear even if reservoirs, by themselves, would locally keep the system in the homogeneous phase. 
This feature is due to the existence of two independent conserved quantities (mass and energy) which determine two macroscopic currents, whose coupling produces a sort of extreme Joule effect~\cite{iubini23}. 

The nonequilibrium condensation mechanism does not manifest itself with a lower number of conserved quantities.
In fact, if only one conservation law is present and both reservoirs are (sub)critical, then the whole system lies in the homogeneous phase.
As a simple argument, let us consider an energy-conserving system in contact with heat baths at inverse temperatures $\beta_L$ and $\beta_R$,
with $\beta_{L,R}>\beta_c$. We assume here that $\beta_c$ is the critical value separating the homogeneous phase (for $\beta\geq\beta_c$) 
from the localized one (for $\beta<\beta_c$). If transport is diffusive, the positivity of heat conductivity $\kappa$ and the Fourier law $\kappa d\beta(x)/dx=$~const imply that steady temperature profiles $\beta(x)$ are always monotonic. Therefore $\beta(x)\geq\beta_c$  in the whole system. 
The same conclusion remains valid even in the presence of space-dependent conductivity $\kappa(x)$.

It is worthy of note that the out-of-equilibrium condensate is qualitatively different from the equilibrium one.
The latter requires to be in the microcanonical ensemble, i.e. to have a perfect conservation of mass and energy. 
For $h>h\ss{C}(a)$ and large $N$, the system can be depicted as a background at critical energy density with the addition of one or more peaks collecting an extra-energy equal to $(h-h\ss{C}(a))N$. 
This feature is reflected in the marginal energy distribution $p(\epsilon)$ on a single site, which displays a high-energy bump as a characteristic signature of the condensed state.
In the out-of-equilibrium condensation, the energy of peaks increases in time (because energy is pumped in the system) until peaks disappear at the system boundaries. Therefore a lattice site $i$ experiences peaks of all energies, so that there is no bump in $p(\epsilon)$.
Furthermore, the order parameter $Y_2(N)$ (participation ratio) is expected not to have a minimum with respect to $N$. 
In this regard,  we recall that the non-monotonic behavior of $Y_2(N)$ in equilibrium conditions provides a practical  and unambiguous criterion for distinguishing finite-size localized states from  delocalized ones, the separation point being precisely the minimum value of $Y_2(N)$~\cite{GIP21}. The monotonic trend of $Y_2(N)$ observed in nonequilibrium conditions prevents such a clear separation.

The above scenario is valid for a wide class of lattice models where masses $c_i$ are  continuous, positive variables
and local energies are $\e_i = F(c_i)$. 
Many of our results have been found analytically and they apply to any convex function $F(c)$. 
Among the main results, we have provided a complete characterization of nonequilibrium profiles and stationary currents which are
valid both in the homogeneous region and in the localized one. 
Moreover, we have derived Onsager coefficients and shown that 
the Seebeck coefficient is finite and nonvanishing on the critical line.
\comment{It will be interesting to analyze the case of a concave $F(c)$, for example $F(c) = c^\alpha$, with $\alpha <1$. 
We argue that the change of variable $x_i = c_{i}^{1/\alpha}$ leads to a new set of conserved quantities, with $\alpha\to 1/\alpha > 1$, but
with mass and energy exhibiting exchanged roles. The resulting physical picture is therefore different from the one studied here and it will deserve future studies.}

At the macroscopic level, localized NESS are profitably described in terms of a non-homogeneous diffusion equation for the excess
energy field $\Delta(x)$ with respect to the local critical energy density $h\ss{C}(x)$, Eq.~(\ref{eq.diffusion}). In particular, we obtain that 
$\Delta(x)$ is such that $\Delta(x)+h\ss{C}(x)$ is a linear function in $x$, see Eq.~(\ref{eq.diffusion2}), which implies that the profiles of
$\Delta(x)$ and $h\ss{C}(x)$ have opposite curvatures. Phenomenologically, the concave shape of $\Delta(x)$ for convex $F(c)$ could be interpreted as a manifestation of a Joule heating effect, whereby the lattice is ``hotter'' in the bulk than at the boundaries. 
It is interesting to note that the source term accounting for energy injection is $I(x)=2d^2 h_C(x)/dx^2$, which 
is constant for the model $F(c)=c^2$. This is the reason why in this model the energy $E^*(t)$ of a peak increases linearly in time,
with a constant slope, independent of the peak, see Fig.~\ref{fig:en_pos_8k}. 
For different models, the $x-$dependence of the source term makes the growth rate of its energy, $dE^*/dt$, dependent on the
peak position, so that $E^*(t)$ is no more a linear function.
In spite of this the overall picture of nonequilibrium condensation is the same:
out-of-equilibrium conditions pump energy in the system inducing the birth of peaks and the increase of their
energy, while they diffuse. Once a peak attains a (sub)critical boundary, it disappears.

The class of models $F(c)=c^\alpha$, with $\alpha>1$, plays a special role because its equilibrium properties have been studied in some details. In particular it has been shown~\cite{Szavits2014_PRL,Gradenigo2021_JSTAT} they are equivalent to models where only the mass $A=\sum_i c_i$ is conserved, but variables $c_i$ are distributed according to $f(c) \sim \exp(-c^{1/\alpha})$. If such models are driven out-of-equilibrium, they do not show any nonequilibrium condensation, as argued here above.

Models where $F(c)$ can be negative  have not been discussed in detail. We limit to observe that a negative, strictly convex function $F(c)$ requires that $\lim_{c\to\infty} |F(c)|/c = 0$, which means that the localization process involves the mass, not the energy.
We remark that this is precisely what happens in the case of the DNLS equation with a saturable nonlinearity~\cite{Samuelsen2013}, a model which can be traced back to the case $F(c)=-\ln (1+c)$.

 Finally, it is interesting to discuss the role of peak diffusion, a key element to obtain parametric linear profiles and to obtain a NESS even when the critical line $h\ss{C}(a)$ is crossed. Let us consider the limiting case in which peak diffusion is suppressed: this scenario can be obtained by constraining the search of the solutions of Eqs.~(\ref{eq.triplet}). Let us discuss explicitly the model $F(c)=c^2$, in which case the new triplet must be chosen in the intersection between a plane and a sphere. The condition of positive masses, $c_i\ge 0$, implies that such intersection may not be a full circle, but the union of three disconnected arcs of circle: this occurs if one of the three masses is much larger than the other two, which is the typical scenario in the condensed region, when the triplet includes a peak. In these cases we may suppress peak diffusion imposing the search of the new triplet in the same arc of circle, a constraint which does not break detailed balance. In fact, equilibrium properties in the presence or in the absence of such additional constraint are the same~\cite{JSP_DNLS}.

 When the system is out-of-equilibrium, driven by boundaries, the above constraint is relevant because enforcing it makes Eqs.~(\ref{eq.triplet2}) no more valid. However, the
 relevance of the constraint depends on temperature, because at $\beta\to+\infty$ the system is perfectly homogeneous and the constraint does not apply: we expect its effects are more and more visible with decreasing $\beta$. In fact,
 the most striking effect of suppressed diffusion occurs in nonequilibrium condensation, hence $\beta(x)<0$, where immobile peaks are unable to discharge their energy, which therefore continues to grow indefinitely: this process prevents a steady state from being reached~\cite{GIP21}.

We conclude by observing that our approach to determine the NESS profiles (spatial and parametric profiles) can be easily extended to a stripe or to a two-dimensional system. 
In this case we have local masses $c_{ij}$ and local energies $\e_{ij} =F(c_{ij})$, where $i=1,\dots, N$ and $j=1,\dots,L$. If sites $(1,j)$ are attached to the reservoir ${\cal R}_L$ and sites $(N,j)$ are attached to the reservoir ${\cal R}_R$, we can apply the same method. We expect to find an analytical solution for the spatial profiles either for small $L$ or for periodic boundary conditions in the direction perpendicular to the currents.

\appendix

\section{Equilibrium properties: microcanonical and grand-canonical descriptions }
\label{app.mc}

In the context of equilibrium statistical mechanics, the condensation phenomenon in models with two conservation laws has been understood from the properties of  the microcanonical partition function (we assume a convex and positive $F(c)$),
\be
\Omega(A,H)=\int_0^\infty \prod_i dc_i\, \delta\left(H-\sum_i F(c_i) \right) \delta\left(A-\sum_i c_i\right)\,.
\label{eq:omega}
\ee
It was shown~\cite{Szavits2014_PRL,Gradenigo2021_JSTAT} that $\Omega(A,H)$ can be conveniently obtained by computing its Laplace transform  $Z(m,\beta)$  with respect to $A=Na$ and $H=Nh$, followed by an inverse Laplace transform. Here $m$ and $\beta$ are real variables conjugate to $A$ and $H$, respectively.
Explicitly, we write the function $Z(m,\beta)$ as
\be
Z(m,\beta)= \int_0^\infty dA\, \int_0^\infty dH\, \Omega(A,H) e^{-\beta H +m A }\,.
\ee
Remarkably, $Z(m,\beta)$ takes the factorized form 
\be
Z(m,\beta)=\left[ \int_0^\infty dc\, e^{-\beta F(c) +m c}    \right]^N
\equiv [z(m,\beta)]^N ,
\ee
with $\beta>0$. The Laplace inversion formula thereby writes
\be
\Omega(A,H)=\int_{m_0-i\infty}^{m_0+i\infty } \frac{dm}{2\pi i}\, \int_{\beta_0-i\infty}^{\beta_0+i\infty} 
\frac{d\beta}{2 \pi i} \,e^{(-m Na+\beta Nh)} Z(m,\beta)\,,
\label{eq:omega2}
\ee
where the parameters $m_0$ and $\beta_0$ define an integration contour free from singularities. In the large $N$ limit, the integral in Eq.~(\ref{eq:omega2})
can be finally evaluated using the saddle-point approximation, solving the following equations
\begin{eqnarray}
    a&=& \frac{\partial \ln\left[z(m,\beta) \right]}{\partial m} =  \frac{\int_0^\infty dc\, c\, e^{(-\beta F(c) +mc)}}{z(m,\beta)} \nonumber\\
    h&=&-\frac{\partial \ln\left[z(m,\beta)\right ]}{\partial \beta} = \frac{\int_0^\infty dc\, F(c)\, e^{(-\beta F(c) +mc)}}
    {z(m,\beta)}
    \label{eq:state}
\end{eqnarray}
From a physical point of view, the above procedure defines the standard grand-canonical approach, where $Z(m,\beta)$ is 
the grand-canonical partition function, $\beta>0$ is the inverse temperature and $m=\beta\mu$ defines implicitly the chemical potential $\mu$.
From the analytic properties of $Z(m,\beta)$ and from Eq.~(\ref{eq:omega2}), it follows that a solution for saddle-point equations~(\ref{eq:state}) exists only for real positive $\beta$ values, or equivalently for $h<h\ss{C}$, see Appendix~\ref{app:b0}. This solution  provides a one-to-one mapping between parameters $(\beta>0,\mu)$ and $(a,h)$.
The following two subsections show the explicit solution for the two relevant limits $\beta\to+\infty$ and $\beta\to 0$.
For $h>h\ss{C}$ (condensed region) Eq.~(\ref{eq:state}) has no solutions and the estimation of the microcanonical partition function $\Omega(a,h)$ 
requires large-deviations techniques~\cite{Szavits2014_PRL}.

\subsection{Limit $\beta\to +\infty$ (ground state)}

In order to obtain a non vanishing partition function, it is necessary to assume $\mu>0$, in which case
\be
z(\mu,\beta) \to \int_0^{c^*} dc e^{\beta(\mu c - F(c))},
\ee
where $\mu c^* = F(c^*)$. The maximum of the exponent is in $c_M$, defined by $\mu=F'(c_M)$, and $a(\beta\to+\infty,\mu) \to c_M$, $h(\beta\to+\infty,\mu) \to F(c_M)$. Therefore, the line $h=F(a)$ is the ground state of the system, as anticipated in the main text by the consideration that it corresponds to a perfectly homogeneous state, $c_i \equiv a$. The above treatment also implies that  the chemical potential on the ground state  is $\mu=F'(a)$.

\subsection{Limit $\beta\to 0$}
\label{app:b0}

In this case the grand-canonical weight is proportional to $e^{\beta\mu c}$, which requires $\beta\mu\equiv m  \to -\gamma$, with $\gamma >0$. We then get $z(\beta\to 0,\mu)=\gamma^{-1}$, $a(\beta\to 0,\mu)=\gamma^{-1}$, and $h(\beta\to 0,\mu)=\gamma \int_0^\infty dc \, F(c) e^{-\gamma c}$. Therefore, the critical line is
\be
h\ss{C}(a) = \int_0^\infty dx F(ax) e^{-x}.
\ee
On this line, the chemical potential diverges according to the relation $\mu\to -1/(a\beta)$.


\section{Calculation of the $A_i$ coefficients }
\label{app:Ai}

Let us consider the system of equations

\bse
    \begin{equation}
    a_1 = a_L
    \end{equation}
    \begin{equation}
    \label{eq:app_A1b}
    a_1 - 4a_2 +2a_3 + a_4 = 0
    \end{equation}
    \begin{equation}
    a_{i-2} + 2a_{i+1} - 6a_{i} + 2 a_{i+1} + a_{i} = 0\quad  3\le i \le N -2
    \end{equation}
    \begin{equation}
    \label{eq:app_A1d}
     a_N - 4a_{N-1} +2a_{N-2} + a_{N-3} = 0
    \end{equation}
    \begin{equation}
    a_N = a_R,  
    \end{equation}
\label{eq:app_A1}

\ese

Since the system is linear with $N$ independent equations and $N$ unknowns $a_i$, if the solution exists, it is unique. We can rewrite the unknowns $a_i$ in the form
\begin{equation}
    a_i = a_L + A_i(a_R - a_L).
    \label{eq:app_A2}
\end{equation}

The system (\ref{eq:app_A1}) is invariant under the exchange $L \to R$, $i \to N+1 -i$, so we can look for a solution with the same symmetry, thus imposing 
\begin{equation}
    A_{N+1-i} = 1 - A_i.
    \label{eq:app_A3}
\end{equation}

Then we can consider an equivalent system of $N$ independent equations obtained either by summing or subtracting the equations which are mapped into each other with the transformation $L \to R$, $i \to N+1 -i$, i.e. the first and last equation, the second and penultimate, the third and third last, and so on. It can be easily demonstrated that all the $a_i$ that satisfy Eq.~(\ref{eq:app_A3}) are already a solution of all the equations obtained summing the equations: for example, summing Eq.~(\ref{eq:app_A1b}) with Eq.~(\ref{eq:app_A1d}), and replacing $a_i$ with (\ref{eq:app_A2}) we get

\begin{equation}
\begin{split}
    &a_1 - 4a_2 + 2 a_3 + a_4 + a_N - 4 a_{N-1} + 2 a_{N-2} + a_{N-3}\\
    =& \Big(A_1 + A_{N} - 4(A_2 + A_{N-1}) + 2 (A_3+A_{N-2}) \\
    & + A_4 + A_{N-3}\Big)(a_R - a_L)  = (1  - 4 + 2 + 1)(a_R - a_L)  = 0
\end{split}
\end{equation}
The equations obtained instead from the differences, rewritten in terms of the $A_i$ coefficients, become  

\bse
\begin{equation}
    A_1 = 0
\end{equation}    
\begin{equation}
    A_1 - 4A_2 +2A_3 + A_4 = 0
\label{eq:app_A4b}    
\end{equation}    
\begin{equation}
    A_{i-2} + 2A_{i+1} - 6A_{i} + 2 A_{i+1} + A_{i} = 0
\label{eq:app_A4c}
\end{equation}
\label{eq:app_A4}
\ese
with $3 \le i \le N/2$ if $N$ is even, and  $3 \le i \le (N-1)/2$ if $N$ is odd. The system of equations to solve has therefore been halved compared to the starting one. From this point on let's focus only on  the even case, with $N = 2m$. 

From Eq.~(\ref{eq:app_A4c}) with $i = m$ we can obtain $A_{m-2}$ as a function of $A_{m-1}$, $A_{m}$, $A_{m+1}$, $A_{m+2}$, 
\begin{equation}
    A_{m-2} = - 2 A_{m-1} + 6 A_{m} - 2 A_{m+1} - A_{m+2}, 
\end{equation}
which replaced in Eq.~(\ref{eq:app_A4c}) with $i = m-1$ gives $A_{m-3}$ as a function of $A_{m-1}$, $A_{m}$, $A_{m+1}$, $A_{m+2}$. Repeating the substitutions iteratively, we can write $A_i$ in the form
\begin{equation}
    A_i = - g_{m-i+2} A_{m-1} + f_{m-i+2} A_{m} - e_{m-i+2} A_{m+1} - d_{m-i+2} A_{m+2}
    \label{eq:app_A5}
\end{equation}
where the coefficients $d_n,e_n,f_n,g_n$ are defined by the recurrence relation
\begin{equation}
    X_{n+4} =  - 2 X_{n+3} + 6 X_{n+2} - 2 X_{n+1} - X_{n},
    \label{eq:app_A11}
\end{equation}
with the following starting values
\begin{equation}
  \begin{split}
      &d_0 =-1, d_1 = 0, d_2 = 0, d_3 = 0 \\
      &e_0 = 0, e_1 =-1, e_2 = 0, e_3 = 0 \\
      &f_0 = 0, f_1 = 0, f_2 = 1, f_3 = 0 \\
      &g_0 = 0, g_1 = 0, g_2 = 0, g_3 =-1 .\\
  \end{split}
  \label{eq:app_A12}
\end{equation}

By using Eq.~(\ref{eq:app_A3}), $A_{m+1} = 1 - A_{m}$, and $A_{m+2} = 1 - A_{m-1}$,  so we can rewrite (\ref{eq:app_A4}) as 

\begin{equation}
    A_i = \alpha_i A_{m-1} + \beta_{i} A_{m} - \gamma_{i},
    \label{eq:app_A6}
\end{equation}
where 
\begin{equation}
\begin{split}
    &\alpha_i = d_{m-i+2} - g_{m-i+2} = (d - g)_{m-i+2}\\
    &\beta_i =  (f + e)_{m-i+2}\\
    &\gamma_i =  (e + d)_{m-i+2}.
\end{split}
\label{eq:app_A9}
\end{equation}  

By imposing $A_1 = 0$, from Eq.~(\ref{eq:app_A6}) with $i=1$ we obtain
\begin{equation}
    A_m = \frac{\gamma_1  - \alpha_1 A_{m-1}}{\beta_1},
\end{equation}
which can be replaced in (\ref{eq:app_A6}) that becomes 
\begin{equation}
    A_i = B_i A_{m-1} - C_i,
    \label{eq:app_A7}
\end{equation}
where 
\begin{equation}
    B_i = \alpha_i  - \alpha_1 \beta_i/\beta_1, \quad     C_i = \gamma_i - \gamma_1 \beta_i/\beta_1.
    \label{eq:app_A8}
\end{equation}

Finally, replacing the $A_i$ from (\ref{eq:app_A7}) in (\ref{eq:app_A4b}), we obtain
\begin{equation}
    A_{m-1} = \frac{4C_2 - 2C_3 - C_4}{4B_2 - 2B_3 - B_4}.
\end{equation}
Once replaced in (\ref{eq:app_A7}), it determines the solution to the system of equations (\ref{eq:app_A4}). The coefficients $A_i$ for $i>m$ are obtained from those with $i\le m$ using Eq.~(\ref{eq:app_A3}). 

The case with $N$ odd, $N=2m+1$, can be analyzed with the same procedure, obtaining Eq.~(\ref{eq:app_A5}) and from this the solution expressed in the form (\ref{eq:app_A7}), with the same definitions of $B_i$ and $C_i$ from (\ref{eq:app_A8}), but with
\begin{equation}
\begin{split}
     &\alpha_i = - g_{m-i+2},\\
     &\beta_i =  (f + d)_{m-i+2},\\ &\gamma_i =  (e/2 + d)_{m-i+2}.     
\end{split}
\label{eq:app_A10}
\end{equation}

Similarly, we can also solve the system of equations associated with the case of the heat baths with ``finite'' efficiency, which correspond to replacing the border condition
\begin{equation}
    A_1 = 0 \to -5 A_1 + A_2 + A_3= 0.  
\end{equation}
Also in this case the solution takes the form (\ref{eq:app_A7}), but with 

\begin{equation}
    A_{m-1} = \frac{C_1 - 4C_2 + 2C_3 + C_4}{B_1 -4B_2 + 2B_3 + B_4},
\end{equation}

\begin{equation}
    B_i = \left(  \alpha_i - \beta_i \frac{\alpha_3 + \alpha_2 - 5 \alpha_1}{\beta_3 + \beta_2 - 5 \beta_1}   \right),
\end{equation}

\begin{equation}
    C_i = \left(  \gamma_i - \beta_i \frac{\gamma_3 + \gamma_2 - 5 \gamma_1}{\beta_3 + \beta_2 - 5 \beta_1}   \right),
\end{equation}
where $\alpha_i, \beta_i, \gamma_i$ are defined as (\ref{eq:app_A8}) with $N$ even and as (\ref{eq:app_A9}) with $N$ odd.

\subsection{Closed-form expressions for  $d, e, f, g$}

The coefficients $d_n, e_n, f_n, g_n$ can be expressed in a closed form solving the recursion relation (\ref{eq:app_A11}). If we consider  $X_n = x^n$, it satisfies Eq.~(\ref{eq:app_A11}) if $x$ is a solution of 

\begin{equation}
    x^4 + 2x^3 - 6 x^2 + 2 x + 1 = 0.
\end{equation}
The roots of this polynomial equation are
\begin{equation}
    x_1 = x_2 = 1, \quad x_3 = -2 + \sqrt{3}\equiv \phi, \quad x_4 = -2 -\sqrt{3} \equiv\psi .
\end{equation}

A general solution $X_n$ of the recursion relation (\ref{eq:app_A11}) is therefore obtained from the roots $x_i$ by considering a linear combination of $x_i^n$. Since the root $x_1 = 1$ occurs 2 times, $n x_1^n = n$ is added to the linear combination, so the general solution of (\ref{eq:app_A11}) is 

\begin{equation}
    X_{n} =  w_1   + w_2 n + w_3 \phi^n + w_4 \psi^n,
\end{equation}
where the coefficients $w_i$ are determined imposing the initial conditions of the recurrence. For $d_n$, $e_n$, $f_n$, $g_n$ this conditions are the ones in (\ref{eq:app_A12}), from which we can finally obtain the closed form expressions

\begin{equation}
  \begin{split}
      &d_n = (6n - 12 + (-12 + 7\sqrt{3})\psi^n - (12 + 7\sqrt{3}) \phi^n)/36 \\
      &e_n =  (6n - 14 + (7 - 4\sqrt{3})\psi^n - (7 + 4\sqrt{3}) \phi^n)/12 \\
      &f_n = (6n - 4 - \phi \psi^n + \psi \phi^n)/12 \\
      &g_n = (6 - 6n  + (-3 + 2\sqrt{3})\psi^n - (3 + 2\sqrt{3}) \phi^n)/36.
  \end{split}
\end{equation}

With this expressions we can therefore obtain a closed form for the coefficients $A_i$. For example, in the case with the heat baths with finite efficiency and even size $N=2m$, $A_1$ can be expressed as 

\begin{equation}
    A_1 = \frac{4 \sqrt{3} - 6 - \phi^{2m} (6 + 4 \sqrt{3})}{ (4\sqrt{3} - 6)m + 5\sqrt{3} - 6 - \phi^{2m} ((6 + 4\sqrt{3})m + 5\sqrt{3} + 6) },
\end{equation}

that for large $N$, since $|\phi| = 2 - \sqrt{3} <1$, can be approximated with $A_1 \approx 1/m = 2/N$.

\section{Onsager coefficients}
\label{app.Onsager}

To obtain the expressions for the Onsager coefficients in the class of models with local energy $\epsilon_i = c_i^\alpha$, it is useful to define  
\begin{equation}
    I(k) = \int_0^\infty  x^k \exp(-\beta x^\alpha + mx) dx.
\end{equation}
With this notation, $I(0)$ is equal to the partition function $z(m, \beta)$, and the $k-$th moment $\langle c^k \rangle$ is

\begin{equation}
    \langle c^k \rangle = \frac{1}{z(\beta, m)} \int_0^\infty  x^k \exp(-\beta x^\alpha + mx) dx = \frac{I(k)}{I(0)}.
\end{equation}
In particular, we have 

\begin{equation}
    a = \frac{I(1)}{I(0)}, \quad  h = \frac{I(\alpha)}{I(0)}.  
\end{equation}

By deriving $I(k)$ with respect to $\beta$ and $m$, we obtain  
\begin{equation}
\begin{split}
    &\partial_m I(k) = \int_0^\infty  x^{k+1} \exp(-\beta x^\alpha + mx) dx = I(k+1)\\
    &\partial_\beta I(k) = - \int_0^\infty x^{k+\alpha} \exp(-\beta x^\alpha + mx) dx = - I(k+\alpha) .
\end{split}
\end{equation}
The Onsager coefficients can be obtained considering the derivatives of $a$ and $h$. For example, to obtain $L_{aa}$, we need to calculate

\begin{equation}
\begin{split}
    &\partial_m a = \partial_m \frac{I(1)}{I(0)} = \frac{I(2)}{I(0)} - \frac{I^2(1)}{I^2(0)},
\end{split}
\end{equation}
and in a similar way we can calculate all the other derivatives to obtain
\begin{equation}
    \begin{split}
       &L_{aa} =  2 \partial_m a = 2\Big(I(2)I(0) - I^2(1)\Big)/I^2(0)\\
       &L_{ah} = - 2 \partial_\beta a 
       = 2\Big(I(\alpha +1)I(0) - I(\alpha)I(1)\Big)/I^2(0)\\
       &L_{ha} = 2 \partial_m h = L_{ah}\\
       &L_{hh} = - 2 \partial\beta h = 2\Big(I(2\alpha)I(0) - I^2(\alpha)\Big)/I^2(0) 
    \end{split}
\end{equation}

If we now want the coefficients $L_{ij}$ on the critical line $h = \Gamma(\alpha + 1)$, lets consider the limit $\beta \to 0$, $m\to -1/a$, in which $I(k)$ becomes 

\begin{equation}
    \lim_{\beta \to 0^+} I(k) = \Gamma(k + 1) a^{k+1}. 
\end{equation}
Therefore, the Onsager coefficients in this limit become 

\begin{equation}
L_{aa}= 2a^2, \quad L_{hh}= 2\big(\Gamma(2 \alpha + 1) - \Gamma^2(\alpha +1)\big)a^{2\alpha }        
\end{equation}

\begin{equation}
    \begin{split}
    L_{ah} = L_{ha} &= 2 (\Gamma(\alpha + 2) - \Gamma(\alpha + 1)) a^{\alpha +1} \\
    & = 2\alpha \Gamma(\alpha +1) a^{\alpha +1}.
    \end{split}
\end{equation}

\section{The diffusion coefficient $D$}
\label{app.D}

The diffusion coefficient $D$ of the models with the evolution rule defined in Sec. \ref{sec.baths} can be obtained making the expression for the probability $P(x,t)$ to find a breather on site $x$ at time $t$ continuous in space and time, so that it can be expressed as a diffusion equation:

\begin{equation}
    \frac{\partial P(x,t)}{\partial t} = D \frac{\partial^2}{\partial x^2} P(x,t).
\end{equation}
In the evolution algorithm, which updates the triplets $T_i =(i-1, i, i+1)$ with $i$ chosen at random, the only triplets that allow the breather's movement are $T_x$, $T_{x-1}$ and $T_{x+1}$, which can be chosen with probability $1/N$. In these cases, after the update the energy peak can be found with probability 1/3 in any one of the sites of the chosen triplet. For example, the breather can reach $x-2$ only when $T_{x-1}$ is chosen, but each one of $T_{x-1}$, $T_{x}$, $T_{x+1}$ allow it to remain in the initial position $x$. The breather also remains in $x$ if the other triplets of which it is not part are chosen.
Therefore, the probability to find the breather in $x$ at time $t+1$ is given by

\begin{equation}
\begin{split}
    &P(x, t+1) = \frac{1}{N}\left(  P(x,t) + \frac{2}{3}(P(x-1,t) + P(x+1,t)) \right.\\
     &+ \left. \frac{1}{3}(P(x-2,t) + P(x+2,t)) \right)  +\frac{N-3}{N} P(x,t). 
\end{split}
\label{app:prob_D}
\end{equation}
Lets Taylor expand (\ref{app:prob_D}) to the first order in $\delta t$, the time increment for a single triplet update: by subtracting $P(x,t)$ from both members and then dividing by $\delta t$ we obtain  
\begin{equation}
\begin{split}
    &\frac{\partial P(x,t)}{\partial t}= \frac{1}{N \delta t } \left( \frac{2}{3} (P(x-1,t) + P(x+1,t)) \right.\\
    &+ \left. \frac{1}{3} (P(x-2,t) + P(x+2,t))  - 2 P(x,t) \right).
\end{split} 
\label{app:prob_D2}
\end{equation}
We can then Taylor expand (\ref{app:prob_D2}) to the second order in space $x$, the first non vanishing order, from which we get    
\begin{equation}
    \frac{\partial P(x,t)}{\partial t} = \frac{2 \delta x^2}{N \delta t}     \frac{\partial^2 P(x,t)}{\partial x^2} = D \frac{\partial^2 P(x,t)}{\partial x^2},
\end{equation}
where $\delta x = 1$ is the distance between two adjacent sites. Time is measured in Monte Carlo steps $\delta \tau$, that corresponds to $N$ moves that increase time by $\delta t$, thus $\delta \tau = N \delta t = 1$. The diffusion coefficient $D$ is then is given by
\begin{equation}
    D = \frac{2 (\delta x)^2}{\delta \tau} = 2.
\end{equation}

\comment{
\begin{acknowledgments}
\end{acknowledgments}
}
\bibliographystyle{spphys}

\begin{thebibliography}{10}
\providecommand{\url}[1]{{#1}}
\providecommand{\urlprefix}{URL }
\expandafter\ifx\csname urlstyle\endcsname\relax
  \providecommand{\doi}[1]{DOI \discretionary{}{}{}#1}\else
  \providecommand{\doi}{DOI \discretionary{}{}{}\begingroup
  \urlstyle{rm}\Url}\fi

\bibitem{eggers99}
J.~Eggers, Physical Review Letters \textbf{83}(25), 5322 (1999)

\bibitem{aggregation_condensation}
C.~Iyer, A.~Das, M.~Barma, Physical Review E \textbf{107}(1), 014122 (2023)

\bibitem{Eisenberg1998}
H.~Eisenberg, Y.~Silberberg, R.~Morandotti, A.~Boyd, J.~Aitchison, Physical
  Review Letters \textbf{81}(16), 3383 (1998)

\bibitem{russell2023localization}
P.S.J. Russell, Y.~Chen, Laser \& Photonics Reviews \textbf{17}(3), 2200570
  (2023)

\bibitem{TS}
A.~Trombettoni, A.~Smerzi, Physical Review Letters \textbf{86}(11), 2353 (2001)

\bibitem{traffic_flow_review}
D.~Chowdhury, L.~Santen, A.~Schadschneider, Physics Reports \textbf{329}(4-6),
  199 (2000)

\bibitem{soh2018jamming}
H.~Soh, M.~Ha, H.~Jeong, Physical Review E \textbf{97}(3), 032120 (2018)

\bibitem{Burda_wealth}
Z.~Burda, D.~Johnston, J.~Jurkiewicz, M.~Kami{\'n}ski, M.A. Nowak, G.~Papp,
  I.~Zahed, Physical Review E \textbf{65}(2), 026102 (2002)

\bibitem{gelation2000}
P.L. Krapivsky, S.~Redner, F.~Leyvraz, Physical review letters \textbf{85}(21),
  4629 (2000)

\bibitem{Pastor2016_SR}
R.~Pastor-Satorras, C.~Castellano, Scientific reports \textbf{6}(1), 1 (2016)

\bibitem{Szavits2014_PRL}
J.~Szavits-Nossan, M.R. Evans, S.N. Majumdar, Physical review letters
  \textbf{112}(2), 020602 (2014)

\bibitem{Levine2005}
E.~Levine, D.~Mukamel, G.~Sch{\"u}tz, Journal of statistical physics
  \textbf{120}(5), 759 (2005)

\bibitem{gotti22}
G.~Gotti, S.~Iubini, P.~Politi, Physical Review E \textbf{106}(5), 054158
  (2022)

\bibitem{iubini23}
S.~Iubini, A.~Politi, P.~Politi, New Journal of Physics \textbf{25}(6), 063020
  (2023)

\bibitem{Iubini2017_Entropy}
S.~Iubini, S.~Lepri, R.~Livi, G.L. Oppo, A.~Politi, Entropy \textbf{19}(9), 445
  (2017)

\bibitem{johansson04}
M.~Johansson, K.{\O}. Rasmussen, Physical Review E \textbf{70}(6), 066610
  (2004)

\bibitem{Samuelsen2013}
M.R. Samuelsen, A.~Khare, A.~Saxena, K.{\O}. Rasmussen, Physical Review E
  \textbf{87}(4), 044901 (2013)

\bibitem{benenti17rev}
G.~Benenti, G.~Casati, K.~Saito, R.S. Whitney, Physics Reports \textbf{694}, 1
  (2017)

\bibitem{Rasmussen2000_PRL}
K.~Rasmussen, T.~Cretegny, P.G. Kevrekidis, N.~Gr{\o}nbech-Jensen, Physical
  review letters \textbf{84}(17), 3740 (2000)

\bibitem{Majumdar2005_PRL}
S.N. Majumdar, M.~Evans, R.~Zia, Physical review letters \textbf{94}(18),
  180601 (2005)

\bibitem{Evans2006_JSP}
M.~Evans, S.N. Majumdar, R.~Zia, Journal of Statistical Physics
  \textbf{123}(2), 357 (2006)

\bibitem{Majumdar2010_LesHouches}
S.~Majumdar, Exact Methods in Low-dimensional Statistical Physics and Quantum
  Computing: Lecture Notes of the Les Houches Summer School: Volume 89, July
  2008 p. 407 (2010)

\bibitem{Gradenigo2021_JSTAT}
G.~Gradenigo, S.~Iubini, R.~Livi, S.N. Majumdar, Journal of Statistical
  Mechanics: Theory and Experiment \textbf{2021}(2), 023201 (2021)

\bibitem{GIP21}
G.~Gotti, S.~Iubini, P.~Politi, Physical Review E \textbf{103}(5), 052133
  (2021)

\bibitem{kevrekidis09}
P.G. Kevrekidis, \emph{The discrete nonlinear Schr{\"o}dinger equation:
  mathematical analysis, numerical computations and physical perspectives},
  vol. 232 (Springer Science \& Business Media, 2009)

\bibitem{arezzo22}
C.~Arezzo, F.~Balducci, R.~Piergallini, A.~Scardicchio, C.~Vanoni, Journal of
  Statistical Physics \textbf{186}(2), 1 (2022)

\bibitem{Iubini2013_NJP}
S.~Iubini, R.~Franzosi, R.~Livi, G.L. Oppo, A.~Politi, New Journal of Physics
  \textbf{15}(2), 023032 (2013)

\bibitem{Gradenigo2021_EPJE}
G.~Gradenigo, S.~Iubini, R.~Livi, S.N. Majumdar, The European Physical Journal
  E \textbf{44}(3), 1 (2021)

\bibitem{evans10}
M.~Evans, S.~Majumdar, I.~Pagonabarraga, E.~Trizac, The Journal of chemical
  physics \textbf{132}(1) (2010)

\bibitem{Szavits2014_JPA}
J.~Szavits-Nossan, M.R. Evans, S.N. Majumdar, Journal of Physics A:
  Mathematical and Theoretical \textbf{47}(45), 455004 (2014)

\bibitem{evans05r}
M.R. Evans, T.~Hanney, Journal of Physics A: Mathematical and General
  \textbf{38}(19), R195 (2005)

\bibitem{JSP_DNLS}
S.~Iubini, A.~Politi, P.~Politi, Journal of Statistical Physics
  \textbf{154}(4), 1057 (2014)

\bibitem{Barre2018_JSTAT}
J.~Barr{\'e}, L.~Mangeolle, Journal of Statistical Mechanics: Theory and
  Experiment \textbf{2018}(4), 043211 (2018)

\bibitem{Lepri_PhysRep}
S.~Lepri, R.~Livi, A.~Politi, Physics reports \textbf{377}(1), 1 (2003)

\bibitem{livi17}
R.~Livi, P.~Politi, \emph{Nonequilibrium statistical physics: a modern
  perspective} (Cambridge University Press, 2017)

\bibitem{baldovin21}
M.~Baldovin, S.~Iubini, Journal of Statistical Mechanics: Theory and Experiment
  \textbf{2021}(5), 053202 (2021)

\bibitem{Baldovin_2021_review}
M.~Baldovin, S.~Iubini, R.~Livi, A.~Vulpiani, Physics Reports \textbf{923}, 1
  (2021)

\end{thebibliography}

\section*{Declarations}

\begin{itemize}

\item Conflict of interest/Competing interests: We declare no competing interests.

\item Data availability: The authors declare that the data supporting the findings of this study are available within the paper.
Row data sets generated during the current study are available from the corresponding author on reasonable request.

\end{itemize}

\end{document}